\documentclass[12pt]{article}
\begin{document}
\newcommand{\beq}{\begin{equation}}
\newcommand{\eeq}{\end{equation}}
\newcommand{\beqa}{\begin{eqnarray}}
\newcommand{\eeqa}{\end{eqnarray}}
\newcommand{\beqar}{\begin{eqnarray*}}
\newcommand{\eeqar}{\end{eqnarray*}}
\newcommand{\al}{\alpha}
\newcommand{\be}{\beta}
\newcommand{\del}{\delta}
\newcommand{\D}{\Delta}
\newcommand{\eps}{\epsilon}
\newcommand{\ga}{\gamma}
\newcommand{\Ga}{\Gamma}
\newcommand{\ka}{\kappa}
\newcommand{\inn}{\!\cdot\!}
\newcommand{\h}{\eta}
\newcommand{\kk}{\varphi}
\newcommand\F{{}_3F_2}
\newcommand{\la}{\lambda}
\newcommand{\La}{\Lambda}
\newcommand{\na}{\nabla}
\newcommand{\Om}{\Omega}
\newcommand{\p}{\phi}
\newcommand{\sig}{\sigma}
\renewcommand{\t}{\theta}
\newcommand{\z}{\zeta}
\newcommand{\ssc}{\scriptscriptstyle}
\newcommand{\eg}{{\it e.g.,}\ }
\newcommand{\ie}{{\it i.e.,}\ }
\newcommand{\labell}[1]{\label{#1}} %{\label{#1}} %
\newcommand{\reef}[1]{(\ref{#1})}
\newcommand{\labels}[1]{\vskip-2ex$_{#1}$\label{#1}} %{\label{#1}} %
\newcommand\prt{\partial}
\newcommand\veps{\varepsilon}
\newcommand\ls{\ell_s}
\newcommand\cF{{\cal F}}
\newcommand\cA{{\cal A}}
\newcommand\cS{{\cal S}}
\newcommand\cH{{\cal H}}
\newcommand\cM{{\cal M}}
\newcommand\cL{{\cal L}}
\newcommand\cG{{\cal G}}
\newcommand\cN{{\cal N}}
\newcommand\cl{{\iota}}
\newcommand\cP{{\cal P}}
\newcommand\cV{{\cal V}}
\newcommand\cg{{\it g}}
\newcommand\cR{{\cal R}}
\newcommand\cB{{\cal B}}
\newcommand\cO{{\cal O}}
\newcommand\tcO{{\tilde {{\cal O}}}}
\newcommand\bz{\bar{z}}
\newcommand\bw{\bar{w}}
\newcommand\hF{\hat{F}}
\newcommand\hA{\hat{A}}
\newcommand\hT{\hat{T}}
\newcommand\htau{\hat{\tau}}
\newcommand\hD{\hat{D}}
\newcommand\hf{\hat{f}}
\newcommand\hg{\hat{g}}
\newcommand\hp{\hat{\phi}}
\newcommand\hh{\hat{h}}
\newcommand\ha{\hat{a}}
\newcommand\hQ{\hat{Q}}
\newcommand\hP{\hat{\Phi}}
\newcommand\hb{\hat{b}}
\newcommand\hc{\hat{c}}
\newcommand\hd{\hat{d}}
\newcommand\hS{\hat{S}}
\newcommand\hX{\hat{X}}
\newcommand\tL{\tilde{\cal L}}
\newcommand\hL{\hat{\cal L}}
\newcommand\tG{{\widetilde G}}
\newcommand\tg{{\widetilde g}}
\newcommand\tphi{{\widetilde \phi}}
\newcommand\tPhi{{\widetilde \Phi}}
\newcommand\te{{\tilde e}}
\newcommand\tk{{\tilde k}}
\newcommand\tf{{\tilde f}}
\newcommand\tF{{\widetilde F}}
\newcommand\tK{{\widetilde K}}
\newcommand\tE{{\widetilde E}}
\newcommand\tpsi{{\tilde \psi}}
\newcommand\tX{{\widetilde X}}
\newcommand\tD{{\widetilde D}}
\newcommand\tO{{\widetilde O}}
\newcommand\tS{{\tilde S}}
\newcommand\tB{{\widetilde B}}
\newcommand\tA{{\widetilde A}}
\newcommand\tT{{\widetilde T}}
\newcommand\tC{{\widetilde C}}
\newcommand\tV{{\widetilde V}}
\newcommand\thF{{\widetilde {\hat {F}}}}
\newcommand\Tr{{\rm Tr}}
\newcommand\tr{{\rm tr}}
\newcommand\STr{{\rm STr}}
\newcommand\Str{{\rm Str}}
\newcommand\M[2]{M^{#1}{}_{#2}}
\parskip 0.3cm
%\begin{document}

%\thispagestyle{empty} \rightline{\small  \hfill IPM/P-2006/xxx}
\vspace*{1cm}

\begin{center}
{\bf \Large  S-duality constraint on higher-derivative couplings}     % at order $\alpha'^3$   }

\vspace*{1cm}

{  Mohammad R. Garousi\footnote{garousi@um.ac.ir} }\\
\vspace*{1cm}
{ Department of Physics, Ferdowsi University of Mashhad,\\ P.O. Box 1436, Mashhad, Iran}
\\
\vspace{2cm}

\end{center}

\begin{abstract}
\baselineskip=18pt

The  Riemann curvature correction to the type II supergravity  at eight-derivative level  in string frame is given    as $ e^{-2\phi}(t_8t_8R^4+\frac{1}{4}\eps_{8}\eps_{8}R^4)$. For constant dilaton, it has been extended in the literature to the S-duality invariant form by extending the dilaton factor in the Einstein frame to the non-holomorphic Eisenstein series. For non-constant dilaton, however, there are various  couplings in the Einstein frame which are not consistent with the   S-duality. By constructing the  tensors $t_{2n}$ from Born-Infeld action, we include   the appropriate Ricci and scalar curvatures as well as the dilaton couplings to make the above action to be consistent with the S-duality.    

\end{abstract}

\setcounter{page}{0}
\setcounter{footnote}{0}
\newpage
%\beqa\frac\eeqa
\section{Introduction  } \label{intro}

Higher-derivative couplings in string theory can be captured by exploring its  wonderful string dualities.     T-duality 
  relates  type IIA superstring theory   at weak (strong) coupling to type IIB superstring theory at weak (strong) coupling \cite{Kikkawa:1984cp}-\cite{Hassan:1999bv}   . At low energy, this duality relates  the type IIA   to  the type IIB  supergravities.   S-duality, on the other hand,    relates the type IIB theory at weak (strong) coupling to the type IIB at strong (weak) coupling \cite{Font:1990gx}-\cite{Hull:1994ys}. At low energy, this is the symmetry of type IIB  supergravity.   The stringy behaviors of the superstring theory which are encoded in the higher-derivative  corrections to these supergravities  should have the same properties. That is, the higher-derivative couplings of type IIB supergravity should be invariant under the S-duality, and the higher-derivative couplings in  type IIA supergravity should be related to the higher-derivative couplings in type IIB supergravity  under the T-duality. These properties may be used as  guiding principles to find the stringy  corrections to  the supergravity. See \cite{Garousi:2009dj}-\cite{ Garousi:2013gea} for related work on higher-derivative couplings of D-brane  action, and \cite{Garousi:2012yr}-\cite{Maxfield:2013wka} for the higher-derivative couplings of the type II supergravities.

The   higher-derivative corrections to the   supergravity   start at the eight-derivative level, and were first found   from the sphere-level four-graviton scattering amplitude   \cite{ Schwarz:1982jn,Gross:1986iv} as well as from the $\sigma$-model beta function approach \cite{Grisaru:1986vi,Freeman:1986zh}. The result in the string frame is  
\beqa
S\supset \frac{\gamma \z(3)}{3.2^7} \int d^{10}x e^{-2\phi} \sqrt{-G}(t_8t_8R^4+\frac{1}{4}\eps_{8}\eps_{8}R^4)\labell{Y0}
\eeqa
where $\gamma=\frac{\alpha'^3}{2^{5}}$ and   $t_8$ is a tensor which is antisymmetric within a pair of indices and is symmetric under exchange of the pair of indices (see equation \reef{t8} for its precise form).  
The couplings given by $t_8t_8R^4$ have nonzero contribution at four-graviton level, so they were found from the sphere-level S-matrix element of four graviton vertex operators \cite{ Schwarz:1982jn,Gross:1986iv}, whereas the couplings given by $\eps_{8}\eps_{8}R^4$ have nonzero contribution at five-graviton level \cite{Zumino:1985dp}. It has been recently shown  this term is consistent with the sphere-level S-matrix element of five graviton vertex operators in the Ramond-Neveu-Schwarz formalism \cite{Garousi:2013tca}.  

The action \reef{Y0} is valid for both type IIA and type IIB theories. In the type IIB case, this action should be extend   to the S-duality invariant form. For constant dilaton, the action \reef{Y0} in the Einstein frame becomes 
\beqa
S\supset \frac{\gamma \z(3)}{3.2^7} \int d^{10}x e^{-3\phi/2} \sqrt{-G}(t_8t_8R^4+\frac{1}{4}\eps_{8}\eps_{8}R^4)\labell{Y1}
\eeqa
The presence of the dilaton factor in this action      indicates that it   needs the genus  and nonperturbative corrections to become S-duality invariant. The   $SL(2,Z)$ invariant form of this action   has been found in  \cite{Green:1997tv} - \cite{Basu:2007ck} to be
\beqa
S\supset\frac{\gamma }{3.2^8 }\int d^{10}x E_{(3/2)}(\tau,\bar{\tau}) \sqrt{-G}(t_8t_8R^4+\frac{1}{4}\eps_{8}\eps_{8}R^4)\labell{Y2}
\eeqa
where $E_{(3/2)}(\tau,\bar{\tau}) $ is the $SL(2,Z)$ invariant non-holomorphic Eisenstein series which has the following weak-expansion \cite{Green:1997tv}:
\beqa
E_{(3/2)}(\tau,\bar{\tau})%&=&\sum_{(m,n)\neq(0,0)}\frac{\tau_2^{3/2}}{|m+n\tau|^{3}}\\
&\!\!\!\!=\!\!\!\!&2\z(3)\tau_2^{3/2}+4\z(2)\tau_2^{-1/2}+8\pi\tau_2^{1/2}\sum_{m\neq 0,n\geq 1}\left|\frac{m}{n}\right| K_{1}(2\pi|mn|\tau_2)e^{2\pi imn\tau_1}\labell{series}
\eeqa
where $\tau=\tau_1+i\tau_2=C_0+ie^{-\phi} $ and $K_1$ is the Bessel function.   The above expansion shows that there are no perturbative corrections beyond the  one-loop level, but there are an infinite number  of D-instanton  corrections.  By explicit calculation, it has been shown in \cite{D'Hoker:2005jc} that there is no two-loop correction to the action \reef{Y2}. The odd-odd coupling $\eps_{8}\eps_{8}R^4$ at one-loop level has been confirmed in \cite{Richards:2008jg,Liu:2013dna} by explicit calculation of  torus-level S-matrix element of  five graviton vertex operators.  In the type IIA case, the sign of the odd-odd term  is minus  at one-loop level, and of course there is no D-instanton corrections. There is also a Chern-Simons term in type IIA case \cite{Vafa:1995fj,Duff:1995wd} in which we are not interested in this paper.

 The non-constant B-field and dilaton couplings at four-field level have been added to   \reef{Y0}    by extending the Riemann curvature to the generalized Riemann curvature at the linear order \cite{Gross:1986mw}\footnote{Note that the normalizations of the dilation  and B-field here  are $\sqrt{2}$ and 2 times the normalization of the dilaton and B-field in \cite{Gross:1986mw}, respectively.},
 \beqa
 \bar{R}_{ab}{}^{cd}&= &R_{ab}{}^{cd}- \eta_{[a}{}^{[c}\phi_{,b]}{}^{d]}+  e^{-\phi/2}H_{ab}{}^{[c,d]}\labell{trans}
\eeqa
where   the bracket notation is $H_{ab}{}^{[c,d]}=\frac{1}{2}(H_{ab}{}^{c,d}-H_{ab}{}^{d,c})$, and comma denotes the partial derivative. 
Using the relation between the Einstein  frame metric and the string frame metric $G_{\mu\nu}=e^{-\phi/2}G^s_{\mu\nu}$, one observes that the dilaton term in above equation is canceled in   transforming the linearized  Riemann curvature from the Einstein frame to the string frame   \cite{Garousi:2012jp}, \ie
\beqa
\bar{R}_{ab cd}&\Longrightarrow&e^{-\phi/2}\cR_{ab cd}
\eeqa
where on the right hand side the metric is in the string frame. In above equation,  $\cR_{ab cd}$ is the following expression 
\beqa
\cR_{ab cd}&=&R_{ab cd}+H_{ab [c,d]}\labell{RH2}
\eeqa
which is the Riemann curvature of the connection with torsion at the linear order, \ie the curvature two-form is
\beqa
\cR^{\alpha\beta}&=&d\tilde{\omega}^{\alpha\beta}\,\,\,;\,\, \tilde{\omega}^{\alpha\beta}=\omega^{\alpha\beta}+\frac{1}{2}H_a{}^{\alpha\beta}dx^a\labell{linear}
\eeqa
The action involving four Neveu-Schwarz-Neveu-Schwarz (NS-NS) fields  at the sphere level then becomes
\beqa
S\supset \frac{\gamma \z(3)}{3.2^7} \int d^{10}x e^{-2\phi} \sqrt{-G}(t_8t_8\cR^4+\frac{1}{4}\eps_{8}\eps_{8}\cR^4)\labell{Y3}
\eeqa
where the metric is in the string frame. The odd-odd coupling $\eps_{8}\eps_{8}\cR^4$ is  total derivative at four-field level.
It has been observed in \cite{Garousi:2012jp, Garousi:2013zca,Liu:2013dna} that the   even-even coupling $t_8t_8\cR^4$   is   invariant under   T-duality. 

The natural nonlinear extension of the generalized Riemann curvature \reef{linear} is
\beqa
\cR^{\alpha\beta}&=&d\tilde{\omega}^{\alpha\beta}+\tilde{\omega}^{\alpha }{}_{\gamma}\wedge \tilde{\omega}^{\gamma\beta } \labell{nonlinear}
\eeqa
which has the following spacetime components:
 \beqa
\cR_{abcd}&=&R_{abcd}+H_{ab[c;d]}-\frac{1}{2}H_{ae[c}H_{|be|d]}\labell{nonlinear2}\\
\cR_{ab}&=&R_{ab}+\frac{1}{2}H_{acb}{}_{;c}-\frac{1}{4}H^2_{ab} \,\,\,;\,\,\,\cR=R-\frac{1}{4}H^2\nonumber
\eeqa
where $\cR_{ab}=\cR_{acb}{}_c$ and the semicolon denotes the covariant derivative. The tours-level coupling of two B-fields and three Riemann curvatures and the coupling of four B-fields and one Riemann curvature in the even-even part have been found in \cite{Richards:2008sa} and shown that they are fully consistent with the corresponding couplings in   $t_8t_8\cR^4$. However, the B-field couplings in the odd-odd sector are not given by $\eps_{8}\eps_{8}\cR^4$.  The one-loop coupling of two B-fields and three Riemann curvatures and the coupling of four B-fields and one Riemann curvature in the odd-odd part have been found in \cite{Peeters:2001ub,Liu:2013dna} and shown that they are not reproduced by the B-field couplings in $\eps_{8}\eps_{8}\cR^4$. One may still extend the curvature in the odd-odd part to the generalized curvature. Then there are  extra couplings in this sector  involving  the field strength $H$ which does not show up in  $\cR$ \cite{Liu:2013dna}. In this paper we   are not interested in fixing such $H$-couplings, so we use only the generalized curvatures throughout this paper.

 Using  the  combination of S- and T-dualities on the action \reef{Y3}, the tensorial structure of various    four-field couplings, including  Ramond-Ramond (R-R) fields,  have been found in \cite{Garousi:2013lja}, and confirmed by the S-matrix calculations in \cite{Garousi:2013tca}. In particular, it has been observed that the Eisenstein series $E_{(3/2)}(\tau,\bar{\tau})$ appears in all couplings, and the extra  dilaton and the axion and their derivatives combine with the other massless fields to become invariant under the $SL(2,R)$ transformation.  In the $SL(2,R)$ form of couplings, one finds no term which has  one dilaton perturbation and three gravitons or   three dilaton perturbations  and one graviton  because it is impossible to write such couplings in $SL(2,R)$ invariant form.  In this study, however, the on-shell relations have been used frequently. 

At the four-field level, it is not hard to study the S-duality  of  various on-shell couplings because there are no massless poles at order $\alpha'^3$. However, at five-field level and higher, there are various massless poles that one should take into account. In general, one expects the S-matrix elements of a field theory which include both massless poles and contact terms to be invariant under the S-dual ward identity \cite{Garousi:2011we}-\cite{Garousi:2012gh}. For example, when transforming the couplings \reef{Y3} to the Einstein frame, one would find non-zero couplings for three dilatons and two gravitons which are not consistent with the S-duality. However, when one combines them with  the  corresponding massless poles, which   produces then the S-matrix element of three dilatons and two gravitons,  one would expect the result to be  zero according to the S-dual ward identity \cite{Garousi:2011we}-\cite{Garousi:2012gh}.  The on-shell action \reef{Y3} then is expected to be consistent with the on-shell S-duality after taking into account the massless poles.

To avoid the massless poles, however,  one may require   the field theory action   to be consistent with the S-duality without using the on-shell relations. Then one would find the    action \reef{Y3} is not consistent with the   S-duality. In particular, when transforming the Riemann curvatures in \reef{Y3} to the Einstein frame, one would find non-zero  couplings between  one dilaton and three  gravitons. These couplings and all other couplings involving the gravitons and odd number of dilaton perturbations are not consistent with the S-duality. In this paper, in order to make this action  to be consistent with the S-duality, we are   going to  include the appropriate Ricci and scalar curvatures as well as the dilaton couplings  in the action \reef{Y3}.  

An outline of the paper is as follows: In section 2 we show when transforming  the action \reef{Y3} to the Einstein frame one finds   couplings involving odd number of dilatons. In this section we include various couplings in the even-even, $t_{2n}t_{2n}$, and the odd-odd, $\eps_n\eps_n$, sectors to remove such undesirable couplings. To construct the even-even couplings, we use  the expansion of the Born-Infeld action to construct the $t_{2n}$ tensors.  We fix the coefficients of the Ricci and the scalar curvature couplings by constraining them to have no coupling of one dilaton and three gravitons in the Einstain frame.    We observe  that, this constraint not only removes the odd number of dilatons, but also it removes all  the couplings between the dilatons and the gravitons in the Einstein frame. In section 3, we then include various couplings between the dilatons and the gravitons in the string frame. In this section we also include the appropriate  couplings of  the Ricci and    scalar curvatures   to make  the dilaton couplings to be  consistent with the S-duality. In section 4, we briefly discuss our results.

\section{$\cR^4$ couplings}

We have seen that for the constant dilaton, the couplings \reef{Y0} can be extended to the S-duality invariant form \reef{Y2}. However, for non-constant dilaton there must be various other couplings to make the acion invariant under the S-duality. In this section we are going to show that in the presence of non-constant dilaton, the S-duality of action \reef{Y0} requires the effective action to have   couplings involving the Ricci and scalar curvatures. So let us  first review the $SL(2,R)$ transformation of various bosonic fields in the supergravity.

Under the $SL(2,R)$ transformation, the  B-field and the R-R two-form transform as doublet  \cite{ Tseytlin:1996it,Green:1996qg}. Since the parameters of the duality are constant, their field strengths, \ie $H= dB$ and $F= dC$,  are  also transform as doublet, 
\beqa
\cH\equiv\pmatrix{H \cr 
F}\rightarrow (\Lambda^{-1})^T \pmatrix{H \cr 
F}\,\,\,;\,\,\,\Lambda=\pmatrix{p&q \cr 
r&s}\in SL(2,R)\labell{2}
\eeqa
The dilaton and the R-R scalar transform non-linearly as $\tau\rightarrow \frac{p\tau+q}{r\tau+s}$. The matrix $\cM$ defined in terms of the dilaton and the R-R scalar, \ie 
 \beqa
 {\cal M}=e^{\phi}\pmatrix{|\tau|^2&C_{0} \cr 
C_{0}&1}\labell{M}
\eeqa
then  transforms  as \cite{Gibbons:1995ap}
\beqa
{\cal M}\rightarrow \Lambda {\cal M}\Lambda ^T
\eeqa
The derivative of this matrix, $\prt\cM$, also transform as above. The Einstein frame metric and the R-R four-form are   invariant under the $SL(2,R)$ transformations. Using the above transformations, one can construct various couplings which are invariant under the $SL(2,R)$ transformations. For example, the coupling $\cH^T\cM\cH$ which has the following components:
\beqa
 \cH^T\cM\cH =e^{-\phi}(1+e^{2\phi}C_0^2)HH+e^{\phi}FF+e^{\phi}C_0(HF+FH)\labell{SHH}
\eeqa
is invariant under the $SL(2,R)$ transformations. The perturbations  of dilaton or axion appears only as $\delta\cM$. One can easily construct the $SL(2,R)$ invariant couplings from two $\cH$s and one dilaton perturbation, \eg
\beqa
 \cH^T\cM_{,h}\cH &=&e^{\phi }(HF+FH) C_{,h} +2 e^{\phi } {H} {H} C_0 C_{,h}+e^{\phi } {F} {F} \phi _{,h}\nonumber\\
&&+e^{\phi } (HF+FH) C_0 \phi _{,h}+e^{\phi } {H} {H} C_0^2 \phi _{,h}-e^{-\phi } {H} {H} \phi _{,h}\labell{HHM}
\eeqa
However, one can   verify that it is impossible to construct $SL(2,R)$ invariant terms  from the gravity and odd number of $\delta\cM$.    In particular, the couplings involving one $\prt^2\phi$ and three gravitons can not be extended to the  $SL(2,R)$ invariant form. %Such couplings are not produced by the S-matrix element of four NS-NS vertex operators either \cite{Garousi:2012yr}. 
Therefore, the effective action in type IIB theory must have no coupling with odd number of dilaton perturbations.     This is the constrain that we are going to impose on the couplings \reef{Y3} to find new couplings involving the Ricci and scalar curvatures.

In order to study the S-duality transformation of the couplings \reef{Y3}, it is  convenient to transform the string frame metric to the Einstein frame metric, \ie $G^s_{\mu\nu}=e^{\phi/2}G_{\mu\nu}$.
For those terms  which have no  derivative of the metric, the transformation  gives only an overall dilaton factor, \eg
\beqa
e^{-2\phi}\sqrt{-G}&\Longrightarrow&e^{\phi/2}\sqrt{-G}
\eeqa
In other cases, there are some extra terms involving the derivative of the dilaton, \eg the transformation of the Riemann curvature      is given by \cite{Garousi:2013gea}
\beqa
R_{\mu\nu\alpha\beta}&\!\!\!\!\Longrightarrow\!\!\!\!&e^{ \phi/2}R_{\mu\nu\alpha\beta}+  e^{ \phi/2}\bigg[G_{[\mu[\beta}\nabla_{\nu]}\prt_{\alpha]}\phi+\frac{1}{4}G_{[\mu[\alpha}\prt_{\nu]}\phi\prt_{\beta]}\phi+\frac{1}{8}G_{[\mu[\beta}G_{\nu]\alpha]}\prt_{\lambda}\phi\prt_{\lambda}\phi\bigg]\labell{StoE}
%\nabla_{\mu}H_{\nu\alpha\beta}&\Longrightarrow&\nabla_{\mu}H_{\nu\alpha\beta}-\frac{3}{4}\bigg[H_{\mu[\nu\alpha}\prt_{\beta]}\phi+H_{\nu\alpha\beta}\prt_{\mu}\phi-G_{\mu[\nu}H_{\alpha\beta]\lambda}\prt^{\lambda}\phi\bigg]\nonumber
\eeqa
where on the right hand side the metric is in the Einstein frame.
Using the above transformations, one can transform the couplings \reef{Y3} to  the Einstein frame  to find various couplings between the  dilatons  and  the gravitons.

Let us start by transforming  the  odd-odd Riemann curvature term in \reef{Y3} to the Einstein frame,
\beqa
\frac{1}{4}e^{-2\phi}\sqrt{-G}\eps_{8}\eps_{8}R^4&\!\!\!\!\!\Longrightarrow \!\!\!\!\!& e^{-3\phi/2}\sqrt{-G}\bigg[\frac{1}{4}\eps_{ \mu_1\cdots\mu_8}\eps_{ \nu_1\cdots\nu_8}R_{\mu_1\mu_2\nu_1\nu_2}R_{\mu_3\mu_4\nu_3\nu_4}R_{\mu_5\mu_6\nu_5\nu_6}R_{\mu_7\mu_8\nu_7\nu_8}\nonumber\\
&&-  \eps_{ \mu_1\cdots\mu_7}\eps_{ \nu_1\cdots\nu_7}\nabla_{\mu_1}\nabla_{\nu_1}\phi R_{\mu_2\mu_3\nu_2\nu_3}R_{\mu_4\mu_5\nu_4\nu_5}R_{\mu_6\mu_7\nu_6\nu_7} \nonumber\\
&&+\frac{3}{2} \eps_{ \mu_1\cdots\mu_6}\eps_{ \nu_1\cdots\nu_6}\nabla_{\mu_1}\nabla_{\nu_1}\phi \nabla_{\mu_2}\nabla_{\nu_2}\phi R_{\mu_3\mu_4\nu_3\nu_4}R_{\mu_5\mu_6\nu_5\nu_6}  \nonumber\\
&&-  \eps_{ \mu_1\cdots\mu_5}\eps_{ \nu_1\cdots\nu_5}\nabla_{\mu_1}\nabla_{\nu_1}\phi \nabla_{\mu_2}\nabla_{\nu_2}\phi \nabla_{\mu_3}\nabla_{\nu_3}\phi R_{\mu_4\mu_5\nu_4\nu_5} \nonumber\\
&&+\frac{1}{4} \eps_{ \mu_1\cdots\mu_4}\eps_{ \nu_1\cdots\nu_4}\nabla_{\mu_1}\nabla_{\nu_1}\phi \nabla_{\mu_2}\nabla_{\nu_2}\phi \nabla_{\mu_3}\nabla_{\nu_3}\phi \nabla_{\mu_4}\nabla_{\nu_4}\phi+\cdots\bigg]\nonumber
\eeqa
where dots refer to the higher order fields which are resulted from the nonlinear dilaton terms in \reef{StoE}. Our notation in  the Levi-Civita tensors $\eps_n\eps_n$ is that $10-n$  indices of the 10-dimensional Levi-Civita tensors are contracted, \eg $\eps_8^{ \mu_1\cdots\mu_8}\eps_8^{ \nu_1\cdots\nu_8}=\eps_{10}^{\mu\nu \mu_1\cdots\mu_8}\eps_{10}^{\mu\nu \nu_1\cdots\nu_8}$. Using the Bianchi identity, one observes that the above couplings which are resulted from the linear dilaton term in \reef{StoE}, are  total derivatives. %Using integration by part, they    can  be written as 
%\beqa
%\frac{1}{8}e^{-2\phi}\sqrt{-G}\eps_{10}\inn\eps_{10}R^4&\!\!\!\!\!\Longrightarrow \!\!\!\!\!& e^{-3\phi/2}\sqrt{-G}\bigg[\frac{1}{8}\eps_{ \mu_1\cdots\mu_8}\eps_{ \nu_1\cdots\nu_8}R_{\mu_1\mu_2\nu_1\nu_2}R_{\mu_3\mu_4\nu_3\nu_4}R_{\mu_5\mu_6\nu_5\nu_6}R_{\mu_7\mu_8\nu_7\nu_8}\nonumber\\
%&&+\frac{3}{4} \eps_{ \mu_1\cdots\mu_7}\eps_{ \nu_1\cdots\nu_7}\nabla_{\mu_1}\phi \nabla_{\nu_1}\phi R_{\mu_2\mu_3\nu_2\nu_3}R_{\mu_4\mu_5\nu_4\nu_5}R_{\mu_6\mu_7\nu_6\nu_7} \nonumber\\
%&&-\frac{9}{8} \eps_{ \mu_1\cdots\mu_6}\eps_{ \nu_1\cdots\nu_6}\nabla_{\mu_1}\phi \nabla_{\nu_1}\phi \nabla_{\mu_2}\nabla_{\nu_2}\phi R_{\mu_3\mu_4\nu_3\nu_4}R_{\mu_5\mu_6\nu_5\nu_6}  \nonumber\\
%&&+\frac{3}{4} \eps_{ \mu_1\cdots\mu_5}\eps_{ \nu_1\cdots\nu_5}\nabla_{\mu_1}\phi\nabla_{\nu_1}\phi \nabla_{\mu_2}\nabla_{\nu_2}\phi \nabla_{\mu_3}\nabla_{\nu_3}\phi R_{\mu_4\mu_5\nu_4\nu_5} \nonumber\\
%&&-\frac{3}{16} \eps_{ \mu_1\cdots\mu_4}\eps_{ \nu_1\cdots\nu_4}\nabla_{\mu_1}\phi \nabla_{\nu_1}\phi \nabla_{\mu_2}\nabla_{\nu_2}\phi \nabla_{\mu_3}\nabla_{\nu_3}\phi \nabla_{\mu_4}\nabla_{\nu_4}\phi+\cdots\bigg]\nonumber
%\eeqa
%Therefore, there are no coupling of one dilaton and three gravitons or three dilatons and one graviton in the odd-odd part. 
However, the higher order terms are not total derivative. In particular, there are couplings between three  dilatons and two curvatures. There are two source for these couplings. One of them is coming from the integration by part of the third term in above equation. The other one is coming from the direct replacement of \reef{StoE} into the odd-odd coupling where appears in the dots in above equation. We have checked that these couplings are not zero.  

The even-even part also produces odd number of dilatons. The transformation of the even-even part to the Einstein frame is 
\beqa
 e^{-2\phi}\sqrt{-G}t_{8}t_{8}R^4&\!\!\!\!\!\Longrightarrow \!\!\!\!\!& e^{-3\phi/2}\sqrt{-G}\bigg[ t_{ \mu_1\cdots\mu_8}t_{ \nu_1\cdots\nu_8}R_{\mu_1\mu_2\nu_1\nu_2}R_{\mu_3\mu_4\nu_3\nu_4}R_{\mu_5\mu_6\nu_5\nu_6}R_{\mu_7\mu_8\nu_7\nu_8}\nonumber\\
&&-4 t_{ \mu_1\cdots\mu_8}t_{ \nu_1\cdots\nu_7\mu_8}\nabla_{\mu_1}\nabla_{\nu_1}\phi R_{\mu_2\mu_3\nu_2\nu_3}R_{\mu_4\mu_5\nu_4\nu_5}R_{\mu_6\mu_7\nu_6\nu_7} \nonumber\\
&&+6 t_{ \mu_1\cdots\mu_8}t_{ \nu_1\cdots\nu_6\mu_7\mu_8}\nabla_{\mu_1}\nabla_{\nu_1}\phi \nabla_{\mu_2}\nabla_{\nu_2}\phi R_{\mu_3\mu_4\nu_3\nu_4}R_{\mu_5\mu_6\nu_5\nu_6}  \nonumber\\
&&-4 t_{ \mu_1\cdots\mu_8}t_{ \nu_1\cdots\nu_5\mu_6\mu_7\mu_8}\nabla_{\mu_1}\nabla_{\nu_1}\phi \nabla_{\mu_2}\nabla_{\nu_2}\phi \nabla_{\mu_3}\nabla_{\nu_3}\phi R_{\mu_4\mu_5\nu_4\nu_5} \nonumber\\
&&+t_{\mu_1\cdots\mu_8}t_{ \nu_1\cdots\nu_4\mu_5\mu_6\mu_7\mu_8}\nabla_{\mu_1}\nabla_{\nu_1}\phi \nabla_{\mu_2}\nabla_{\nu_2}\phi \nabla_{\mu_3}\nabla_{\nu_3}\phi \nabla_{\mu_4}\nabla_{\nu_4}\phi+\cdots\bigg]\nonumber
\eeqa
where dots refer to the higher order fields which are resulted from the nonlinear dilaton terms in \reef{StoE}.   Since the $t_8$ tensor is not totally antisymmetric, the above dilaton couplings are not total derivative terms. In particular, the couplings of one dilaton and three curvatures or three dilatons and one curvature are not zero. %Using the on-shell relations,  one finds that these couplings are zero. However, we are interested in finding an action which is manifestly invariant under the S-duality. 
Therefore, as in the odd-odd sector, there are  couplings which have odd number of dilatons.
% in higher order terms which are not zero even on-shell. We have checked explicitly the couplings of three dilatons and two gravitons in above equation and found that they are not zero even after using onshell relation. We have also combined such couplings with the corresponding couplings in the odd-odd secor, and found that they do not vanish after using on-shell relations. 

Since there are couplings of odd number of  dilatons, one concludes that the couplings in \reef{Y3} are not consistent with the S-duality for non-constant dilaton field. To remedy this failure one may add some new four-curvature couplings to \reef{Y3}. Such couplings can not be  captured by the S-matrix calculations, so they must involve the Ricci and/or scalar curvatures.   The transformation of these couplings to the Einstein frame should then chancel the above couplings which have odd number of dilatons. Since the couplings of one dilaton and three curvatures are not zero, we  impose the above condition on these coupling.

%The generalized Riemann curvature \reef{nonlinear2} gives the following generalized Ricci and scalar curvatures:
%\beqa
%\cR_{ab}=R_{ab}+\frac{1}{2}H_{acb;c}-\frac{1}{4}H^2_{ab} &;&\cR=R-\frac{1}{4}H^2
%\eeqa
%In the absence of the B-field and dilaton, the Ricci curvature satisfies on-shell relation $R_{ab}=0$. In the presence of the B-field this equation extends to  $R_{ab}-\frac{1}{4}H^2_{ab}=\cR_{ab}=0$. So it is consistent with on-shell physics to extend the Ricci curvature to generalized Ricci curvature. One the other hand, the scalar curvature satisfies on-shell relation $R=0$  in the absence of B-field and dilaton, whereas, in the presence of the B-field this extend to   $R-\frac{1}{12}H^2=0$, \ie $\cR\neq 0$.  Therefore, it is not consistent with on-shell physics to extend the scalar curvature to the generalized scalar curvature.

To construct  various  couplings between four curvatures, we need to define some tensors that contract appropriately with the indices of the four curvatures. Assuming   the Kawai-Lewellen-Tye relation \cite{Kawai:1985xq}  is holed for the closed string couplings, the tensors should be square of some lower rank tensors. We call these lower rank tensor,   the open string tensors  and the square of them,   the   closed string tensors. For example, to have the coupling of three Riemann curvatures and one Ricci curvature, we need an open string tensor of rank  seven. The only possibility for the closed string tensor with rank 14 is $ \eps_{ \mu_1\cdots\mu_7}\eps_{ \nu_1\cdots\nu_7}$. The coupling is then $\eps_7\eps_7\cR^4$. Similarly, one may use the  10-dimensional Levi-Civita tensor to construct the closed string tensors with lower rank, \ie $\eps_6\eps_6\cR^4$, $\eps_5\eps_5\cR^4$, $\eps_4\eps_4\cR^4$ which have no scalar curvatures. The couplings which have one scalar curvature are $\cR\eps_6\eps_6\cR^3$, $\cR\eps_5\eps_5\cR^3$, $\cR\eps_4\eps_4\cR^3$, and $\cR\eps_3\eps_3\cR^3$. The couplings which have two scalar curvatures are  $\cR^2\eps_4\eps_4\cR^2$
 and  $\cR^2\eps_2\eps_2\cR^2$. And there is one coupling which has  four scalar curvatures, \ie $\cR^4$. The rank of the odd-odd tensor $\eps_n\eps_n$ dictates how many of the curvatures in the above couplings are the Ricci curvature, so we don't need to specify how many of the  curvature are the Riemann and how many of them are the Ricci curvature.  For example  the coupling $\eps_6\eps_6\cR^3$ has three Riemann curvatures, and the coupling $\eps_6\eps_6\cR^4$ has two Ricci and two Riemann curvatures. These odd-odd couplings can be expanded using the relation
\beqa
 \eps_{ \mu_1\cdots \mu_n}\eps^{ \nu_1\cdots\nu_n}&=&-n!\delta^{\nu_1}_{[\mu_1}\cdots \delta^{\nu_n}_{\mu_n]}\labell{expand}
\eeqa
 Using the expansion form of each coupling, one finds that there is no coupling for odd number of B-field strength, which is consistent with parity.

For the open string tensor with rank even, however, there are other possibilities. Since the open string tensors  should  appear also in the effective action of D-brane, we construct these tensors from expanding the   Born-Infeld Lagrangian. So consider the following expansion:
\beqa
 \sqrt{-\det(\eta +M )}&\!\!\!\!\!=\!\!\!\!\!&1+\frac{1}{2}\Tr(M)-\frac{1}{4}\Tr(M^2)+\frac{1}{8}(\Tr(M))^2+\frac{1}{6}\Tr(M^3)+\frac{1}{48}(\Tr(M))^3\nonumber\\
&&-\frac{1}{8}\Tr(M)\Tr(M^2)-\frac{1}{32}(\Tr(M))^2\Tr(M^2)+\frac{1}{32}(\Tr(M^2))^2\nonumber\\
&&+\frac{1}{12}\Tr(M)\Tr(M^3)-\frac{1}{8}\Tr(M^4)+\frac{1}{384}(\Tr(M))^4+\cdots\labell{BI}
\eeqa
 where $M$ is an arbitrary  matrix.  When one deals with the couplings which involve only the generalized Riemann curvature, the matrix $M$ should be chosen to be  antisymmetric. In other cases, this matrix has both the symmetric and the antisymmetric parts.  
 
 The coupling of four arbitrary antisymmetric matrices $M^1,\,\cdots, M^4$   gives the tensor $t_8$ which was first defined in \cite{Schwarz:1982jn} by other means, \ie 
\beqa
&&\frac{1}{8}t_{8}M^1 M^2 M^3 M^4 =-\bigg[\Tr( M^1M^2M^3M^4)+\Tr (M^1M^3M^2M^4)+\Tr (M^1M^3M^4M^2)\bigg]\nonumber\\
&&\quad +\frac{1}{4}\bigg[\Tr (M^1M^2)\Tr (M^3M^4)+\Tr (M^1M^3)\Tr (M^2M^4)+\Tr (M^1M^4)\Tr (M^2M^3)\bigg]\labell{t8}
\eeqa
where we have added the factor $1/8$ to have the same normalization for $t_8$ as in \cite{Schwarz:1982jn}. Our prescription for constructing the above tensor is the following:  we have first replaced the matrix $M$ in \reef{BI} with $M=M^1+M^2+M^3+M^4$ and kept the terms which have $M^1M^2M^3M^4$. Then we have replaced   each structure with  average of all independent contractions with identical wight, \ie $3\Tr( M^1M^2M^3M^4)$ is replaced by the expression in the first line above.  Note that there are 6 non-cyclic permutations for this term, however, only three of them are independent.  

Writing similar expression for four other antisymmetric matrices $\tilde{M}^1,\,\cdots, \tilde{M}^4$, and writing the Riemann curvature as $\cR_{\mu\nu\alpha\beta}=M^i_{\mu\nu}\tilde{M}^i_{\alpha\beta}$, one finds the coupling $t_8t_8\cR^4$ which has the following expansion \cite{Garousi:2013zca}:
 \beqa
 t_8t_8\cR^4&=& 3.2^7  \bigg[\cR_{h k m n} \cR_{k r n p}\cR_{r s m q}\cR_{h s p q}+\frac{1}{2} \cR_{h k m n} \cR_{k r n p}\cR_{r s p q} \cR_{h s m q}\nonumber\\&&-\frac{1}{4} \cR_{h k m n} \cR_{h k n s} \cR_{p q m r} \cR_{p q r s}+\frac{1}{8} \cR_{h k m n} \cR_{h k r s} \cR_{p q n r} \cR_{p q m s}\nonumber\\&&+\frac{1}{4} \cR_{h k m n} \cR_{k r m n}\cR_{r s p q} \cR_{h s p q}+\frac{1}{8} \cR_{h k m n} \cR_{k r p q}\cR_{r s m n} \cR_{h s p q}\nonumber\\&&+\frac{1}{16} \cR_{h k m n} \cR_{h k p q} \cR_{r s m n} \cR_{r s p q}+\frac{1}{32} \cR_{h k m n}\cR_{h k m n} \cR_{r s p q}\cR_{r s p q}\bigg]\labell{t8t8}
 \eeqa 
where   the antisymmetry property of the first and the second pairs of the indices in the generalized Riemann curvature has been used in above expansion. The above couplings do not produce odd number of B-field strength \cite{Garousi:2013zca}.

Now to construct the tensor which   contracts with two generalized Riemann  and two generalized Ricci curvatures, we first write the Ricci curvature as $\cR_{\mu\nu}=L^i_{\mu}\tilde{L}^i_{\nu}$. Then we write the matrix $M$ to be $M=\frac{1}{2}(L^1L^2+L^2L^1)+M^3+M^4$ where the antisymmetric matrices $M^3$ and $M^4$ correspond to the Riemann curvatures. Replacing it in \reef{BI}, keeping the terms which have $L^1L^2M^3M^4$ and replacing each structure with  average of all independent contractions, one finds  our definition of tensor $t_6$ which is
\beqa
\frac{1}{8}t_6L^1L^2M^3M^4&=&\frac{1}{2}\bigg[L^1\inn M^4M^3\inn L^2+L^1\inn M^3M^4\inn L^2\bigg]-\frac{1}{4}L^1\inn L^2\Tr(M^3M^4)\labell{LL}
\eeqa
Writing similar expression for $t_6\tilde{L}^1\tilde{L}^2\tilde{M}^3\tilde{M}^4$, and using the relations  $\cR_{\mu\nu\alpha\beta}=M^i_{\mu\nu}\tilde{M}^i_{\alpha\beta}$ for $i=3,4$ and $\cR_{\mu\nu}=L^i_{\mu}\tilde{L}^i_{\nu}$ for $i=1,2$, one finds the coupling $t_6t_6\cR^4$ which has the following expansion:
\beqa
t_6t_6\cR^4&=&64\bigg[\frac{1}{2} \cR_{hm} \cR_{kn} \cR_{hpnr} \cR_{kpmr}+\frac{1}{2} \cR_{hm} \cR_{kn} \cR_{hpmr} \cR_{kpnr}\nonumber\\
&&\quad-\frac{1}{2} \cR_{hm} \cR_{km} \cR_{hpnr} \cR_{kpnr}+\frac{1}{16} \cR_{hm}^2 \cR_{kpnr}^2\bigg]\labell{t6t6RR}
\eeqa
We have checked that the above couplings do not produce odd number of B-field strength, which is consistent with parity.

To construct the tensor which is contracted with four generalized Ricci curvatures, we write the matrix $M=\frac{1}{2}(L^1L^2+L^2L^1)+\frac{1}{2}(L^3L^4+L^4L^3)$. Performing the same steps as before, one finds  our definition of tensor $t_4$ which is
\beqa
\frac{1}{8}t_4L^1L^2L^3L^4&=&\frac{1}{4}\bigg[-L^1\inn L^4 L^2\inn L^3-L^1\inn L^3L^2\inn L^4+L^1\inn L^2L^3\inn L^4\bigg]\labell{LLLL}
\eeqa
Writing similar expression for $t_4\tilde{L}^1\tilde{L}^2\tilde{L}^3\tilde{L}^4$, and using the relation  $\cR_{\mu\nu}=L^i_{\mu}\tilde{L}^i_{\nu}$ for $i=1,2,3,4$, one finds the coupling $t_4t_4\cR^4$ which has the following expansion:
\beqa
t_4t_4\cR^4&=&64\bigg[-\frac{1}{8} \cR_{hm} \cR_{hn} \cR_{km} \cR_{kn}+\frac{3}{16} \cR_{hm}^2 \cR_{kn}^2\bigg]\labell{t4t4}
\eeqa
We have checked that the above couplings do not produce odd number of B-field strength.

The above even-even couplings are the only four curvature couplings which have Ricci and Riemann curvatures. However, there are even-even couplings which involve scalar curvature. These couplings can also easily be constructed. The even-even couplings involving one scalar curvature must have three Riemann curvatures which can be constructed by inserting the antisymmetric matrix $M=M^1+M^2+M^3$ into \reef{BI}. It  gives the following result:
\beqa
\frac{1}{8}t'_6M^1M^2M^3&=&\Tr(M^1M^2M^3)
\eeqa
Witting similar expression for $t'_6\tilde{M}^1\tilde{M}^2\tilde{M}^3$, one finds
\beqa
\cR t'_6t'_6\cR^3&=&64\cR \cR_{h km n} \cR_{h pm r} \cR_{k pn r}\labell{RRR}
%\nonumber\\\cR t'_4t'_4\cR^3&=&\cR \cR_{h m} \cR_{k n} \cR_{h km n}
\eeqa
The   couplings $\cR^2\cR_{hkmn}^2$ and $\cR^2\cR_{hk}^2$  can also be constructed which have two scalar curvatures. However, these two couplings  are not independent of the coupling that we have considered in the odd-odd sector.

Having found all independent couplings of four curvatures, we now add them to the couplings \reef{Y3}  with unknown coefficients, \ie
\beqa
\cL&
 \supset
&\frac{\gamma\z(3)}{3.2^7}e^{-2\phi}\sqrt{-G}\bigg[t_8t_8\cR^4+a_1 t_6t_6\cR^4+a_2 t_4t_4\cR^4+a_3\cR t'_6t'_6\cR^3 +a_4\cR^4\labell{tot}\\
&&\qquad\qquad+\frac{1}{4}\eps_{8}\eps_8\cR^4+b_1\eps_7\eps_7\cR^4+b_2\eps_6\eps_6\cR^4+b_{3}\eps_5\eps_5\cR^4+b_{4}\eps_4\eps_4\cR^4+b_5\cR\eps_6\eps_6\cR^3\nonumber\\
&&\qquad\qquad+b_6\cR\eps_5\eps_5\cR^3+b_7\cR\eps_4\eps_4\cR^3+b_8\cR\eps_3\eps_3\cR^3+ b_9\cR^2\eps_4\eps_4\cR^2+b_{10}\cR^2\eps_2\eps_2\cR^2\bigg]\nonumber
\eeqa
 Using the identity \reef{expand} for expanding the odd-odd couplings, and using the expansion form of the couplings in the even-even sector, one may rewrite the above couplings in terms of contractions of four curvatures. Then using   \reef{StoE}, one transforms the resulting  couplings to the Einstein frame and imposes the condition that there must be no coupling of one dilaton and three curvatures in the Einstein frame.  We have found that the even-even couplings together do not satisfy this constraint. They should be combined with the odd-odd couplings to satisfy the S-duality constraint. This constraint fixes uniquely all the unknown coefficient to be the following: 
\beqa
&&a_1=-3,\,\,a_2=\frac{3}{4},\,\, a_3=0,\,\,a_4=-\frac{539}{18}\labell{consa}\\
 &&b_1=-2,\,b_2=9,\,b_3=-24,\,b_4= 29,\,b_5= \frac{2}{3},\nonumber\\
&&b_6=-8,\,b_7= 36,\,b_8= -60,\, b_9= 2,\,b_{10}= -28\nonumber
\eeqa
These constants depend on the spacetime dimension which we have evaluated them for $D=10$. Note that the four Riemann curvature couplings $(t_8t_8+\frac{1}{4}\eps_8\eps_8)\cR^4$ which have been used to fix these numbers are valid only in 10 dimensions. One may use the  identity $\eps_2\eps_2\cR^2=-\cR^2+t_2t_2\cR^2$ where $t_2t_2\cR^2=\cR_{\mu\nu}\cR_{\mu\nu}$, to write  the last term    in \reef{tot} in terms of $\cR^2t_2t_2\cR^2$. In that case the coefficient of $\cR^4$ would be $a_4=-\frac{35}{18}$.

We have found the above numbers by imposing the condition that the couplings of one dilaton and three curvatures in the Einstain frame is zero. However, using these numbers, one finds the Lagrangian \reef{tot}    produces no coupling between gravity   and  odd number of dilatons. Even more, it produces   nigher the  coupling between dilatons nor the couplings between dilatons and gravity     when transforming it to the Einstein frame.   Such   couplings    must  then be included in the action  as new couplings in the string frame. Moreover, the above action produces couplings between dilatons and B-fields. In particular the couplings between two dilatons and two B-fields are not zero. However, these couplings are not consistent with the corresponding S-matrix element. On the other hand, if we consider only the first term in \reef{tot}, then the    couplings of two dilatons and two B-fields are reproduced exactly by the S-matrix element \cite{Garousi:2013tca}. The reason for this strange point is that the Ricci curvature couplings in \reef{tot} which are zero on-shell, produce nonzero  dilaton couplings when transforming them to the Einstein frame, \ie 
\beqa
R_{\mu\nu}\Longrightarrow R_{\mu\nu}-2\phi_{;\mu\nu}-\frac{1}{4}\nabla^2\phi G_{\mu\nu}+\frac{1}{2}\prt_{\mu}\phi\prt_{\nu}\phi-\frac{1}{2}(\prt\phi)^2G_{\mu\nu}
\eeqa
while the left hand side is zero on-shell, the right hand side is not zero.  This inconsistency with the S-matrix again indicates that there must be another coupling between two dilatons and two B-fields in the string frame. We will study these couplings in the next section.

After transforming the Lagrangian \reef{tot} to the Einstein frame, it produces the couplings between one dilaton, one graviton and two B-fields which can be written in $SL(2,R)$ invariant form \reef{HHM}. On the other hand, it has been shown in \cite{Garousi:2013tca} that such couplings which  are produced  by the first term in \reef{tot}, are exactly reproduced by the corresponding S-matrix element. This indicates that the corresponding   couplings from all other terms in \reef{tot} must be zero on-shell. We have checked this explicitly and found positive answer.

We have found the gravity couplings in \reef{tot} by imposing the condition that there is no odd number of dilaton in the Einstein frame. The B-field couplings in \reef{tot} can be transformed to the Einstein frame using the transformation
\beqa
\nabla_{\mu}H_{\nu\alpha\beta}&\Longrightarrow&\nabla_{\mu}H_{\nu\alpha\beta}-\frac{3}{4}\bigg[H_{\mu[\nu\alpha}\prt_{\beta]}\phi+H_{\nu\alpha\beta}\prt_{\mu}\phi-G_{\mu[\nu}H_{\alpha\beta]\lambda}\prt_{\lambda}\phi\bigg]\nonumber
\eeqa 
However, the S-duality does not constraint that there must be no odd number of dilatons because one can construct $SL(2,R)$ scalars from odd number of dilatons and even number of B-fields. In fact, since the gravity and the B-field behave totally differently under the S-duality, one does not expect the S-duality invariant action to be in terms of the generalized curvatures in which the graviton and B-field appear symmetrically at the linear order. As a result,  one expects,  not all the B-field couplings are given by the generalized curvatures  in \reef{tot}. There are other couplings involving the field strength H in which we are not interested in this paper.

Before ending this section, let us make a comment about the presence of the Ricci and scalar curvatures in the higher-derivative action. If one is interested only in the gravity,  then such couplings can be absorbed  by field redefinition of metric into the Hilbert-Einstein action \cite{Gross:1986iv}. So they can be simply dropped from the action. However, 
 if one is interested in  the couplings of all components of the supergravity multiplet, as we are, then removing the Ricci and scalar curvatures by field redefinition produces new couplings between other components of the supergravity multiplet. %So it is not a priori clear that it is economical to remove the Ricci and scalar curvatures before finding couplings of all other components.
 
\section{Dilaton couplings }

We have seen that the string frame action \reef{tot}   produces no couplings between the  dilatons and  the gravity  or among the   dilatons when transforming it to  the Einstein frame.   However, the S-matrix element of  NS-NS vertex operators produces such couplings in the Einstein frame. Therefore, there must be  dilaton  couplings in the S-duality invariant action in the string frame. To find such couplings, we use the construction of the even-even couplings from the Born-Infeld action as in the previous section. 

\subsection{$(\nabla^2\phi)^2\cR^2$ and $(\nabla^2\phi)^4$couplings}

The S-matrix element of  four NS-NS vertex operators produces   couplings between two dilatons and two gravitons, and  also   couplings between two dilatons and two B-fields. We have seen that the former couplings are not produced by the  couplings \reef{tot} at all and a part of the latter coupling are produced by \reef{tot}. One can verify that in the odd-odd sector the couplings of two Riemann curvatures and two dilatons  at order $\alpha'^3$, \ie $\eps_6\eps_6(\nabla^2\phi)^2\cR^2$,  are  total derivative terms at the four-field level, hence, we have to construct the on-shell string theory couplings in the  even-even sector.

To find the couplings between two   dilatons and two Riemann curvatures at eight derivative level in the even-even sector, we   write  $\nabla_{\mu}\nabla_{\nu}\phi=L_{\mu}^i\tilde{L}_{\nu}^i$ for $i=1,2$. Then doing the same steps that lead to the coupling \reef{t6t6RR}, one finds    the coupling $t_6t_6(\nabla^2\phi)^2\cR^2$ which has the following expansion:
\beqa
t_6t_6(\nabla^2\phi)^2\cR^2&=&64\bigg[\frac{1}{2} \phi_{;hm} \phi_{;kn} \cR_{hpnr} \cR_{kpmr}+\frac{1}{2} \phi_{;hm} \phi_{;kn} \cR_{hpmr} \cR_{kpnr}\nonumber\\
&&\quad-\frac{1}{2} \phi_{;hm} \phi_{;km} \cR_{hpnr} \cR_{kpnr}+\frac{1}{16} \phi_{;hm}^2 \cR_{kpnr}^2\bigg]\labell{t6t6pp}
\eeqa
Interestingly, the above couplings are exactly the couplings of two dilatons and two Riemann curvatures which have been found in \cite{Garousi:2013lja}  from the combination of the S- and T- dualities on the couplings \reef{Y3}, and have been verified by  the S-matrix calculations in \cite{Garousi:2013tca} (see eq.(25) in \cite{Garousi:2013lja}). The above equation   includes also  the couplings of two dilatons and two B-fields which are not consistent with the couplings that have been found in \cite{Garousi:2013lja}. In particular,   no coupling with the structure $\phi_{;hm}^2 H_{kp[n,r]}^2$ has been found in \cite{Garousi:2013lja} (see eq.(32) in \cite{Garousi:2013lja}). The reason for this discrepancy is that as we mentioned before, the Lagrangian \reef{tot}  produces also the on-shell couplings between two dilatons and two B-fields. We have checked that  the sum of these two contributions is reproduced exactly by the corresponding S-matrix element. 

The even-even coupling \reef{t6t6pp} however   produces the couplings of three dilatons and one Riemann curvature   when transforming it to the Einstein frame which is  not consistent with the S-duality. So as in the previous section we have to add the generalized Ricci and/or  scalar curvatures in the string frame to remove such undesirable couplings. There are two such couplings in the even-even sector and five couplings in the odd-odd sector. The couplings in the odd-odd sector are  $\eps_5\eps_5(\nabla^2\phi)^2\cR^2$, $\eps_4\eps_4(\nabla^2\phi)^2\cR^2$, $\cR\eps_3\eps_3(\nabla^2\phi)^2\cR$ and $\cR^2\eps_2\eps_2(\nabla^2\phi)^2$. One may also consider the generalized Riemann curvature coupling $\eps_6\eps_6(\nabla^2\phi)^2\cR^2$ which is a total derivative term at the four-field level.

To construct the couplings in the even-even sector, we write $\nabla_{\mu}\nabla_{\nu}\phi=L_{\mu}^i\tilde{L}_{\nu}^i$ for $i=1,2$  and  $\cR_{\mu\nu}=L^i_{\mu}\tilde{L}^i_{\nu}$ for $i=3,4$. Then doing the same steps that lead to the coupling \reef{t4t4}, one finds  the coupling $t_4t_4(\nabla^2\phi)^2\cR^2$ which has the following expansion:
\beqa
t_4t_4(\nabla^2\phi)^2\cR^2=4\bigg[2 \phi_{;hm} \phi_{;kn} \cR_{hn} \cR_{km}+2 \phi_{;hm} \phi_{;kn} \cR_{hm} \cR_{kn}-4 \phi_{;hm} \phi_{;hn} \cR_{km} \cR_{kn}+\phi_{;hm}^2 \cR_{kn}^2\bigg]\nonumber
\eeqa
The other  coupling  in this sector is   $\cR^2 t_2t_2(\nabla^2\phi)^2=\cR^2\nabla_{\mu}\nabla_{\nu}\phi\nabla_{\mu}\nabla_{\nu}\phi$. Note that the coupling  $\cR^2 (\nabla^2\phi)^2$, is not independent of the couplings $\cR^2\eps_2\eps_2(\nabla^2\phi)^2$ and $\cR^2 t_2t_2(\nabla^2\phi)^2$.

Using the normalization of the coupling \reef{t6t6pp} which is consistent with \reef{Y3}, and adding the other couplings with unknown coefficients, \ie
\beqa
\cL&\!\!\!\!\!\!\!\supset\!\!\!\!\!\!\!&\frac{\gamma\z(3)e^{-2\phi}}{2^7}\sqrt{-G}\bigg[t_6t_6(\nabla^2\phi)^2\cR^2\!+\!\alpha_1 t_4t_4(\nabla^2\phi)^2\cR^2\!+\!\alpha_2\cR^2 t_2t_2(\nabla^2\phi)^2\!+\!\beta_1\eps_6\eps_6(\nabla^2\phi)^2\cR^2\nonumber\\
&&+\beta_2 \eps_5\eps_5(\nabla^2\phi)^2\cR^2+\beta_3 \eps_4\eps_4(\nabla^2\phi)^2\cR^2+\beta_4 \cR\eps_3\eps_3(\nabla^2\phi)^2\cR+\beta_5 \cR^2\eps_2\eps_2(\nabla^2\phi)^2\bigg]\labell{del}
\eeqa
one can find the coefficients by imposing the condition that there is no coupling of three dilatons and one curvature when transforming them to the Einstein frame. This fixes uniquely  the constants to be
\beqa
&&\alpha_1=-\frac{1}{2},\, \alpha_2=\frac{1}{9}\\
&&\beta_1=\frac{7}{15},\,\beta_2=-\frac{4}{3},\,\beta_3=0,\,\beta_4=2,\,\beta_5=-1\nonumber
\eeqa
Using the above numbers, we have checked that the couplings \reef{del} do not produce four dilaton couplings when transforming them to the Einstein frame. So the four dilaton couplings should be added in the string frame as new couplings.

To construct the tensor which is contracted with four dilatons in the even-even sector, we write $\nabla_{\mu}\nabla_{\nu}\phi=L_{\mu}^i\tilde{L}_{\nu}^i$ for $i=1,2,3,4$. Then using the expansion \reef{LLLL}, one finds   the coupling $t_4t_4(\nabla^2\phi)^4$ which has the following expansion:
\beqa
t_4t_4(\nabla^2\phi)^4&=&64\bigg[-\frac{1}{8} \phi_{;hm} \phi_{;hn} \phi_{;km} \phi_{;kn}+\frac{3}{16} \phi_{;hm}^2 \phi_{;kn}^2\bigg]\labell{t4t40}
\eeqa
Using the on-shell relation $\phi_{;hm}^2 \phi_{;kn}^2=2\phi_{;hm} \phi_{;hn} \phi_{;km} \phi_{;kn}$ \cite{Garousi:2012jp}, one finds the above couplings are exactly reproduced by S-matrix element of four dilaton vertex operators \cite{Garousi:2013tca}. 
Using the normalization   which is consistent with \reef{Y3}, the dilaton coupling is  
\beqa
\cL&\supset&\frac{\gamma\z(3)}{2^5}e^{-2\phi}\sqrt{-G}\,t_4t_4(\nabla^2\phi)^4 \labell{4del}
\eeqa
One may also consider the odd-odd coupling $\eps_4\eps_4(\nabla^2\phi)^4$ which is the same as \reef{t4t40} using the on-shell relations. The S-duality constraint can not relate  these two couplings.

The couplings \reef{del} and \reef{4del} produce non-zero couplings for five  dilatons when transforming them to the Einstein frame. This is resulted from the transformation of the second   derivative of the dilaton in these couplings to the Einstein frame, \ie 
\beqa
\nabla_{\mu}\prt_{\nu}\phi &\Longrightarrow&\nabla_{\mu}\prt_{\nu}\phi -\frac{1}{2} \prt_{\mu}\phi\prt_{\nu}\phi+\frac{1}{4}G_{\mu\nu}\prt_{\alpha}\phi\prt_{\alpha}\phi \,,\labell{rel}
\eeqa
The nonlinear term produces odd number of dilatons in transforming the couplings in \reef{del} and \reef{4del} to the Einstein frame which is not consistent with the S-duality. To avoid this undesirable property, we have to add some couplings in the string frame which involve higher order  of the dilaton. To this end, we define the operator $\bar{\nabla}^2_{\mu\nu}$ in the string frame to be
\beqa
\bar{\nabla}^2_{\mu\nu}\phi&\equiv&\nabla_{\mu}\prt_{\nu}\phi +\frac{1}{2} \prt_{\mu}\phi\prt_{\nu}\phi-\frac{1}{4}G_{\mu\nu}\prt_{\alpha}\phi\prt_{\alpha}\phi 
\eeqa
Then under the transformation from the string frame to the Einstein frame, it  obviously     transforms as   
\beqa
\bar{\nabla}^2_{\mu\nu}\phi&\Longrightarrow&\nabla_{\mu}\prt_{\nu}\phi
\eeqa
Using this operator, we extend the couplings \reef{del} and \reef{4del} to
\beqa
\cL&\!\!\!\!\!\supset\!\!\!\!\!&\frac{\gamma\z(3)}{2^7}e^{-2\phi}\sqrt{-G}\bigg[t_6t_6(\bar{\nabla}^2\phi)^2\cR^2-\frac{1}{2} t_4t_4(\bar{\nabla}^2\phi)^2\cR^2+\frac{1}{9}\cR^2 t_2t_2(\bar{\nabla}^2\phi)^2+\frac{7}{15}\eps_6\eps_6(\bar{\nabla}^2\phi)^2\cR^2\nonumber\\
&&-\frac{4}{3} \eps_5\eps_5(\bar{\nabla}^2\phi)^2\cR^2 +2 \cR\eps_3\eps_3(\bar{\nabla}^2\phi)^2\cR- \cR^2\eps_2\eps_2(\bar{\nabla}^2\phi)^2+4t_4t_4(\bar{\nabla}^2\phi)^4\bigg]\labell{del2}
\eeqa
which includes higher order of dilaton couplings in the string frame. This Lagrangian does not produce odd number of dilatons when transforming it to the Einstein frame, so it is consistent with the S-duality. 

Following \cite{Garousi:2013lja}, one can use the combination of S- and T-dualities to find the couplings of two dilatons and two curvatures from the couplings of two B-fields and two curvatures in \reef{tot}. We have found that the $(\prt H)^2R^2$ couplings in the first term of \reef{tot} produces exactly the $(\prt^2\phi)^2R^2$ coupling in the first term in above equation, however, the $(\prt H)^2R^2$ couplings in other terms in \reef{tot} do not produce the other $(\prt^2\phi)^2R^2$ couplings in \reef{del2}. This indicates that, as we have anticipated before, not all the B-field couplings in a manifestly S-duality invariant theory are given by the generalized curvatures in which the gravity and B-field appear symmetrically at the linear order. This constraint may be used to find the complete B-field couplings in \reef{tot} which have structure $(\prt H)^2R^2$.

The   action \reef{del2} for the gravity part is complete and  can be extended to the S-duality invariant form by including new terms. In the Einstein frame the overall dilaton factor   becomes $e^{-3\phi/2}$ which is extended to the  non-holomorphic Eisenstein series $E_{3/2}(\tau, \bar{\tau})$ after including the one loop result and the nonperturbative effects. The second derivatives of  the dilatons, on the other hand, are extended to the $SL(2,R)$ invariant form
\beqa
 \Tr[\cM_{;hk}\cM^{-1}_{;mn} ]&=& -2\phi_{;hk}\phi_{;mn}+2\phi _{,h} \phi _{,k} \phi _{,m} \phi _{,n}+2 e^{2 \phi } \bigg[-C_{;hk}C_{;mn}-C_{,k} C_{;m n} \phi _{,h}\nonumber\\
&&-C_{,h} C_{;m n} \phi _{,k}-C_{;h k} C_{,n} \phi _{,m}-C_{;h k} C_{,m} \phi _{,n}+C_{,m} C_{,n} \phi _{;h k}+C_{,h} C_{,k} \phi _{;m n}\nonumber\\
&&+C_{,m} C_{,n} \phi _{,h} \phi _{,k}-C_{,k} C_{,n} \phi _{,h} \phi _{,m}-C_{,h} C_{,n} \phi _{,k} \phi _{,m}-C_{,k} C_{,m} \phi _{,h} \phi _{,n}\nonumber\\
&&-C_{,h} C_{,m} \phi _{,k} \phi _{,n}+C_{,h} C_{,k} \phi _{,m} \phi _{,n} \bigg]\labell{FF}
\eeqa 
after including the couplings of four dilatons and all other couplings involving  the R-R scalar field. The R-R scalar couplings should be related by the combination of S- and T-dualities to the B-field couplings in \reef{tot} which have structures $H(\prt H)\prt^2\phi R^2$ and $H^2(\prt\phi)^2R^2$ in the Einstein frame. This gives another constraint on the B-field couplings in which we are not interested in this paper. 

The S-duality invariant form of the last term in \reef{del2} should have two $SL(2,R)$ scalars \reef{FF}.  However, there are ambiguities in choosing which   pair of dilatons should appear in the first $SL(2,R)$ scalar. The R-R couplings in the S-duality invariant form then gives information about the B-field couplings with structure $H^2(\prt\phi)^2(\prt^2\phi)^2$, $H^2(\prt\phi)^6$, $(\prt H)^2(\prt^2\phi)^2$, $(\prt H)^2(\prt\phi)^4$, $H(\prt H)(\prt^2\phi)^2\prt\phi$, $H(\prt H)(\prt\phi)^5$, $H^2(\prt H)^2(\prt\phi)^2$, $H^3(\prt H)(\prt \phi)^3$, $H(\prt H)^3\prt\phi$,  $H^4(\prt\phi)^4$ and $(\prt H)^4$.

%The gravity part in \reef{del2} then can be easily extended to the $SL(2,Z)$ invariant form.  The two B-fields part should be combined with the couplings of two dilatons and two B-fields which are resulted from transforming the couplings \reef{tot} to the Einstein frame. Then they can be extended to the $SL(2,R)$ invariant from \reef{SHH} after including the appropriate R-R couplings. The Lagrangian \reef{del2} contains also the couplings of four B-fields and two dilatons which have no unique $SL(2,Z)$ invariant form.

\subsection{$(\prt\phi)^2\cR^3$ couplings}

We now consider the couplings in the string frame which have $(\prt\phi)^2$ and three curvatures. In the even-even sector, the couplings can be $(\prt\phi)^2t_6t_6\cR^3$, $t_6t_6(\prt\phi)^2\cR^3$, $t_4t_4(\prt\phi)^2\cR^3$ and $(\prt\phi)^2\cR^3$.  The first coupling is similar to the coupling \reef{RRR}. Using $\prt_{\mu}\phi\prt_{\nu}\phi=L^1_{\mu}\tilde{L}_{\nu}^1$ and $\cR_{\mu\nu}=L^2_{\mu}\tilde{L}_{\nu}^2$, then the expansion \reef{LL} leads to the following expansion for $t_6t_6(\prt\phi)^2\cR^3$:
\beqa
t_6t_6(\prt\phi)^2\cR^3&=&64\bigg[\frac{1}{2} \phi _{,h} \phi _{,m} \cR_{kn} \cR_{hpnr} \cR_{kpmr}+\frac{1}{2} \phi _{,h} \phi _{,m} \cR_{kn} \cR_{hpmr} \cR_{kpnr}\nonumber\\
&&\quad-\frac{1}{2} \phi _{,h} \phi _{,m} \cR_{hn} \cR_{kpmr} \cR_{kpnr}+\frac{1}{16} \phi _{,h} \phi _{,m} \cR_{hm} \cR_{kpnr}^2\bigg]
\eeqa
Using $\prt_{\mu}\phi\prt_{\nu}\phi=L^1_{\mu}\tilde{L}_{\nu}^1$    and $\cR_{\mu\nu}=L^i_{\mu}\tilde{L}_{\nu}^i$ for $i=2,3,4$, then the expansion \reef{LLLL} leads to the following expansion for $t_4t_4(\prt\phi)^2\cR^3$:
\beqa
t_4t_4(\prt\phi)^2\cR^3&=&64\bigg[-\frac{1}{8} \phi _{,h} \phi _{,m} \cR_{hn} \cR_{km} \cR_{kn}+\frac{3}{16} \phi _{,h} \phi _{,m} \cR_{hm} \cR_{kn}^2\bigg]
\eeqa
%One may find the coefficient of these three even-even couplings by using duality transformations on the couplings \reef{tot}. However,  we are going to find these coefficients by requiring that there is no odd number of dilatons in the Einstein frame.
In the odd-odd sector, there are   couplings with structure $\eps\eps(\prt\phi)^2\cR^3$, $\cR\eps\eps(\prt\phi)^2\cR^2$ and  $\cR^2\eps\eps(\prt\phi)^2\cR$ in which the dilatons contract with the Levi-Civita tensors, and the couplings    $(\prt\phi)^2\eps\eps\cR^3$ and $(\prt\phi)^2\cR\eps\eps\cR^2$. 

Consider all the above couplings with unknown coefficients, \ie 
\beqa
\cL&\supset&e^{-2\phi}\sqrt{-G}\bigg[m_1\eps_7\eps_7(\prt\phi)^2\cR^3+m_2\eps_6\eps_6 (\prt\phi)^2\cR^3+m_3\eps_5\eps_5(\prt\phi)^2\cR^3+m_4\eps_4\eps_4(\prt\phi)^2\cR^3\nonumber\\
&&+m_5\cR\eps_5\eps_5 (\prt\phi)^2\cR^2+m_6\cR\eps_4\eps_4(\prt\phi)^2\cR^2+m_7\cR\eps_3\eps_3(\prt\phi)^2\cR^2+m_8\cR^2\eps_3\eps_3 (\prt\phi)^2\cR\nonumber\\
&&+m_9\cR^2\eps_2\eps_2(\prt\phi)^2\cR+m_{10}(\prt\phi)^2\cR\eps_4\eps_4\cR^2+m_{11}(\prt\phi)^2\cR\eps_3\eps_3\cR^2+m_{12}(\prt\phi)^2\cR\eps_2\eps_2\cR^2\nonumber\\
&&+m_{13} (\prt\phi)^2\eps_6\eps_6\cR^3+m_{14}(\prt\phi)^2\eps_5\eps_5\cR^3+m_{15}(\prt\phi)^2\eps_4\eps_4\cR^3+m_{16} (\prt\phi)^2\eps_3\eps_3\cR^3\nonumber\\
&&+m_{17} t_6t_6(\prt\phi)^2\cR^3+m_{18}t_4t_4(\prt\phi)^2\cR^3 +m_{19}(\prt\phi)^2\cR^3+m_{20} (\prt\phi)^2t_6t_6\cR^3\bigg]\labell{2del}
\eeqa
where  $m_1,\cdots, m_{20}$ are the unknown coefficients. One may  try to fix the coefficients by combining the above couplings with \reef{del} and then transforming  them to the Einstein frame. In that case one would find it is impossible to constrain them to satisfy the S-duality condition. So the couplings in \reef{del} must be separately extended to satisfy the  S-duality  condition, as we have done by extending $\nabla_{\mu}\nabla_{\nu}\phi$ to $\bar{\nabla}^2_{\mu\nu}\phi$, and the above couplings   should separately satisfy this constraint.

Unlike the previous cases that the S-duality constraint connects all terms together, in this case the constraint does not connect all the above terms. In fact the S-duality  fixes $m_{17}=m_{18}=0$, and gives five multiples. Two of them, \ie  
\beqa
&&m_8 \bigg[ \cR^2\eps_3\eps_3 (\prt\phi)^2\cR-4\cR^2\eps_2\eps_2(\prt\phi)^2\cR-2(\prt\phi)^2\cR^3 \bigg]\,=\,0\nonumber\\
&&m_{11}\bigg[(\prt\phi)^2\cR\eps_3\eps_3\cR^2-4(\prt\phi)^2\cR\eps_2\eps_2\cR^2-2(\prt\phi)^2\cR^3\bigg]\,=\,0\nonumber
\eeqa
  which can easily be verified using the expansion of the Levi-Civita tensors \reef{expand}. In fact the above relations show that not all the couplings that we have considered in \reef{2del} were independent. The other three multiplets are 
\beqa
\cL&\supset& e^{-2\phi}\sqrt{-G}\bigg[m_1\bigg(\eps_7\eps_7(\prt\phi)^2\cR^3-6\eps_6\eps_6 (\prt\phi)^2\cR^3+15\eps_5\eps_5(\prt\phi)^2\cR^3-15\eps_4\eps_4(\prt\phi)^2\cR^3\nonumber\\
&&\qquad\qquad+\frac{5}{3}\cR\eps_5\eps_5 (\prt\phi)^2\cR^2-10\cR\eps_4\eps_4(\prt\phi)^2\cR^2+\frac{35}{2}\cR\eps_3\eps_3(\prt\phi)^2\cR^2 +\frac{70}{27}(\prt\phi)^2\cR^3\bigg)\nonumber 
\\&&\qquad\qquad
+m_{13} \bigg(\frac{5}{2}(\prt\phi)^2\cR\eps_4\eps_4\cR^2-35(\prt\phi)^2\cR\eps_2\eps_2\cR^2+ (\prt\phi)^2\eps_6\eps_6\cR^3\nonumber\\
&&\qquad\qquad-\frac{15}{2}(\prt\phi)^2\eps_5\eps_5\cR^3+\frac{45}{2}(\prt\phi)^2\eps_4\eps_4\cR^3-\frac{105}{4} (\prt\phi)^2\eps_3\eps_3\cR^3-\frac{770}{27}(\prt\phi)^2\cR^3\bigg)\nonumber
\\&&\qquad\qquad+m_{20} \bigg(-\frac{7}{6}(\prt\phi)^2\cR\eps_4\eps_4\cR^2+\frac{85}{3}(\prt\phi)^2\cR\eps_2\eps_2\cR^2+\frac{3}{2}(\prt\phi)^2\eps_5\eps_5\cR^3\nonumber\\
&&\qquad\qquad-\frac{9}{2}(\prt\phi)^2\eps_4\eps_4\cR^3-\frac{11}{4} (\prt\phi)^2\eps_3\eps_3\cR^3+\frac{1438}{81}(\prt\phi)^2\cR^3+  (\prt\phi)^2t_6t_6\cR^3\bigg)\bigg]\labell{2del5}
\eeqa
 As in the previous cases, each multiplet contain one term which is not zero on-shell. In the first, the second and in the third multiples they are $\eps_7\eps_7(\prt\phi)^2\cR^3$, $(\prt\phi)^2\eps_6\eps_6\cR^3$, and $(\prt\phi)^2t_6t_6\cR^3$, respectively. As a result the coefficients of the above three multiplets, \ie $m_1,\, m_{13},\, m_{20}$,  may be found from the S-matrix element of three gravitons and two dilatons in which we are not interested in this paper. 

However, the following duality argument shows that the constant $m_{20}$ may be zero:    As we have seen in the Introduction section, there are evidences to believe that the B-field couplings in the even-even sector appear only through the generalized Riemann curvature.   Replacing the generalized curvature \reef{nonlinear2} into the even-even coupling $t_8t_8\cR^4$ and using the expansion \reef{t8t8},  one finds various  couplings between two $H$s and three Riemann curvatures. However, it is easy to check that   the  two $H$s  in these coupling    contract only once. The $SL(2,R)$ transformation then indicates that the $(F^{(3)})^2R^3$ couplings in the even-even sector has no term in which two $F^{(3)}$s   contract at least twice. As a result, the T-duality indicates    that   there is no coupling $(F^{(1)})^2R^3$ in the even-even sector \cite{Garousi:2013lja}. The $SL(2,R)$ symmetry then indicates that there is no coupling with structure $(\prt\phi)^3R^3$ where $R$ is the Riemann curvature tensor. %Therefore, $m_{20}=0$.   %This is also consistent with our speculation that the couplings involving odd number of the Riemann curvatures in the even-even sector are zero.

The   Lagrangian \reef{2del5} does not produce couplings between gravity  and odd number of dilatons when transforming it to the Einstein frame. So it is consistent with the S-duality.    The overall dilaton factor for three gravity part  becomes $e^{-3\phi/2}$ which is extended to the  non-holomorphic Eisenstein series $E_{3/2}(\tau, \bar{\tau})$ after including the one loop result and the nonperturbative effects. The first derivatives of  the dilatons, on the other hand, are extended to the $SL(2,R)$ invariant form
\beqa
 \Tr[\cM_{,h}\cM^{-1}_{,m} ]&=&-2e^{2\phi}C_{,h}C_{,m}-2\phi_{,h}\phi_{,m}\labell{FF2}
\eeqa 
after including the first derivatives of the R-R scalar field.  The R-R scalar couplings then give information about the B-field couplings in \reef{tot} which have structure $H^2R^3$.

%The couplings of two dilatons, two B-fields and one graviton in \reef{2del5} can also be easily extended to the S-duality invariant from because it has one $\Tr[\delta \cM\delta\cM^{-1}]$ and one $\cH^T\cM\cH$. The couplings of three dilatons  and two B-fields which have contributions from \reef{2del5} and \reef{tot}, however, have ambiguity in writing them in the $SL(2,Z)$ invariant from because the couplings contain $\Tr[\delta \cM\delta\cM^{-1}]$ and  $\cH^T\delta\cM\cH$. The ambiguity is in choosing which dilaton perturbation should appear in  $\cH^T\delta\cM\cH$.

\subsection{$(\prt\phi)^4\cR^2$ couplings}
 
 We now construct the couplings in the string frame which have $(\prt\phi)^4$ and two curvatures. In the even-even sector, there are the couplings $t_6t_6(\prt\phi)^4\cR^2$   and $(\prt\phi)^4\cR^2$. To construct the first coupling, one considers $\nabla_{\mu}\phi\nabla_{\nu}\phi=L^i_{\mu}\tilde{L}^i_{\nu}$ for $i=1,2$. Then using \reef{LL}, one finds the following expansion for the coupling $t_6t_6(\prt\phi)^4\cR^2$:
 \beqa
t_6t_6(\prt\phi)^4\cR^2&=&64\bigg[ \phi_{,h} \phi_{,m}\phi_{,k} \phi_{,n}\cR_{hpnr} \cR_{kpmr} \nonumber\\
&&\quad-\frac{1}{2} (\prt\phi)^2\phi_{,h} \phi_{,k} \cR_{hpnr} \cR_{kpnr}+\frac{1}{16} (\prt\phi)^4 \cR_{kpnr}^2\bigg]\labell{t6t6pppp}
\eeqa
   In the odd-odd sector there are  couplings with structure $ (\prt\phi)^2\eps\eps(\prt\phi)^2\cR^2$,  and  couplings with structure $ (\prt\phi)^4\eps\eps\cR^2$. Note that there is no coupling with structure $\eps\eps(\prt\phi)^4\cR^2$ because two $\prt\phi$ must be contracted with one of the Levi-Civita tensor which is zero. Adding these terms with unknown coefficients and imposing the condition that there must be no coupling between five dilatons and one curvature, one finds the coefficient of the even-even term \reef{t6t6pppp} to be zero. This is resulted from the fact that the terms in the odd-odd sector have no term like the first term in \reef{t6t6pppp}. The other terms  group into two multiplets, \ie
\beqa
\cL&\supset& e^{-2\phi}\sqrt{-G}\bigg[a\bigg((\prt\phi)^2\eps_5\eps_5(\prt\phi)^2\cR^2-6(\prt\phi)^2\eps_4\eps_4(\prt\phi)^2\cR^2\nonumber\\
&&\qquad\qquad\qquad+\frac{21}{2}(\prt\phi)^2\eps_3\eps_3(\prt\phi)^2\cR^2+\frac{21}{9}(\prt\phi)^4\cR^2\bigg)\nonumber\\
&&\qquad\quad\quad+b\bigg(  (\prt\phi)^4\eps_4\eps_4\cR^2- 14(\prt\phi)^4\eps_2\eps_2\cR^2-\frac{91}{9}(\prt\phi)^4\cR^2\bigg)\bigg]\labell{ab}
\eeqa
Here again the first terms of the   multiplets are not zero on-shell, so the coefficients $a,b$ may be found by the S-matrix element of four dilatons and two gravitons.  

Note that the even-even coupling \reef{t6t6pppp} in the string frame is not consistent with the S-duality. However, there is such coupling in the Einstein frame which is coming from  extending the couplings \reef{del2} to the $SL(2,Z)$ invariant form using the $SL(2,R)$ invariant expression  \reef{FF}.   

  In the Einstein frame the overall dilaton factor in \reef{ab} for the gravity part becomes $e^{-3\phi/2}$ which is extended to $E_{(3/2)}(\tau,\bar{\tau})$. Using the $SL(2,R)$ invariant expression \reef{FF2}, the dilatons can be extended to $\Tr[\prt\cM\prt\cM^{-1}]\Tr[\prt\cM\prt\cM^{-1}]$ after including the   R-R scalars. However, there are ambiguities in choosing which   pair of dilatons should appear in the first term. The couplings of four R-R scalars and two curvatures which are unambitious, are related by the dualities to the couplings in \reef{tot} which have structure $H^4R^2$ in which we are not interested.

%The couplings of two B-fields and four dilatons have contributions from \reef{ab}, \reef{2del5} and \reef{tot}. Their $SL(2,Z)$ invariant form has structure $E_{(3/2)}\Tr[\prt\cM\prt\cM^{-1}]\Tr[\prt\cM\prt\cM^{-1}]\cH^T\cM\cH$. The dilatons again have the same ambiguities as in the gravity part.

\subsection{$(\nabla^2\phi)^2(\prt\phi)^2\cR$ couplings}

We now consider the couplings which have $(\nabla^2\phi)^2(\prt\phi)^2$ and one curvature. In the even-even sector the coupling is $t_4t_4(\nabla^2\phi)^2(\prt\phi)^2\cR$. Writing $\nabla_{\mu} \nabla_{\nu}\phi=L^i_{\mu}\tilde{L}^i_{\nu}$ for $i=1,2$, $ \nabla_{\mu}\phi\nabla_{\nu}\phi=L^3_{\mu}\tilde{L}^3_{\nu}$ and $\cR_{\mu\nu}=L^4_{\mu}\tilde{L}_{\nu}^4$, then the expansion \reef{LLLL} leads to the following expression:
\beqa
t_4t_4(\nabla^2\phi)^2(\prt\phi)^2\cR&=&4\bigg[2 \phi _{,k} \phi _{;h m} \phi _{,n} \phi _{;k n} \cR_{hm}-4 \phi _{,k} \phi _{;h m} \phi _{;k m} \phi _{,n} \cR_{hn}\nonumber\\
&&\quad+2 \phi _{,k} \phi _{,m} \phi _{;h m} \phi _{;k n} \cR_{hn}+\phi _{,k} \phi _{;h m}^2 \phi _{,n} \cR_{kn}\bigg]
\eeqa
In the odd-odd sector, one has the couplings with structure $\eps\eps(\nabla^2\phi)^2 (\prt\phi)^2\cR$,  $(\prt\phi)^2\eps\eps(\nabla^2\phi)^2\cR$   and  $\cR\eps\eps(\nabla^2\phi)^2(\prt\phi)^2$. In this case the couplings which satisfy the S-duality constraint are the following:
\beqa
\cL\supset c e^{-2\phi}\sqrt{-G}\bigg[\eps_5\eps_5(\bar{\nabla}^2\phi)^2 (\prt\phi)^2\cR-3 \eps_4\eps_4(\bar{\nabla}^2\phi)^2 (\prt\phi)^2\cR+\frac{7}{6}\cR\eps_3\eps_3(\bar{\nabla}^2\phi)^2(\prt\phi)^2\bigg]\labell{c}
\eeqa
where we have also used the replacement $\nabla_{\mu}\nabla_{\nu}\phi\rightarrow \bar{\nabla}^2_{\mu\nu}\phi$ . In above equation,  the first term is not zero on-shell, so the overall constant $c$ may be calculated from the S-matrix element of four dilatons and one graviton in which we are not interested in this paper.

The couplings in \reef{c} are consistent with the S-duality. In the Einstein frame  the overall dilaton factor is $e^{-3\phi/2}$. The S-duality invariant form of the couplings has the structure $E_{(3/2)}\Tr[\prt^2\cM\prt^2\cM^{-1}]\Tr[\prt\cM\prt\cM^{-1}]R$ which has ambiguity in the dilaton terms.

Finally, the couplings  which have $(\prt\phi)^6$ and one curvature are $t_4t_4(\prt\phi)^6\cR$ and  $(\prt\phi)^6\cR$ in the even-even sector, and the coupling $(\prt\phi)^4\eps_2\eps_2(\prt\phi)^2\cR$ in the odd-odd sector. Note that  coupling $(\prt\phi)^4\eps_3\eps_3(\prt\phi)^2\cR$  is not independent of the other couplings.  Writing $ \nabla_{\mu}\phi\nabla_{\nu}\phi=L^i_{\mu}\tilde{L}^i_{\nu}$ for $i=1,2,3$ and $\cR_{\mu\nu}=L^4_{\mu}\tilde{L}_{\nu}^4$, then the expansion \reef{LLLL} leads to the following expression:
\beqa
t_4t_4(\prt\phi)^6\cR&=&4(\prt\phi)^4\phi_{,h}\phi_{,k}\cR_{hk}
\eeqa
So this term is not independent of the other two couplings either. One can easily check that  it is impossible to constrain the couplings $(\prt\phi)^6\cR$ and $(\prt\phi)^4\eps_2\eps_2(\prt\phi)^2\cR$  to be consistent with the S-duality. So their coefficients must be zero. It is also consistent with our observation that each S-duality multiplet should contain one term which is non-zero on-shell. Note that the symmetries of the Riemann curvature do not permit us to construct the  coupling between six dilatons and one Riemann curvature. 

The coupling with structure  $(\prt\phi)^8$ already appears in the $SL(2,Z)$ invariant form of the last term in \reef{del2}. However, that term is not related to the coupling of eight R-R scalars. The eight dilaton couplings which are related to eight R-R scalars or to the $H^8$ appear in the S-duality invariant  structure  $E_{(3/2)}(\Tr[\prt\cM\prt\cM^{-1}])^4$.

 \section{Discussion}

In this paper we have shown that   in order to have a manifestly S-duality invariant action for the dilaton couplings, the $SL(2,Z)$ invariant   action \reef{Y2} should be extended to the following action:
\beqa
S&\supset&\frac{\gamma }{3.2^8 }\int d^{10}x E_{(3/2)}(\tau,\bar{\tau})\sqrt{-G}\bigg[t_8t_8R^4-3 t_6t_6R^4+\frac{3}{4} t_4t_4R^4 -28R^2t_2t_2R^2 -\frac{35}{18}R^4\nonumber\\
&&\qquad\qquad+\frac{1}{4}\eps_{8}\eps_8R^4-2\eps_7\eps_7R^4+9\eps_6\eps_6R^4-24\eps_5\eps_5R^4+29\eps_4\eps_4R^4+\frac{2}{3}R\eps_6\eps_6R^3\nonumber\\
&&\qquad\qquad-8R\eps_5\eps_5R^3+36R\eps_4\eps_4R^3-60 R\eps_3\eps_3R^3+ 2R^2\eps_4\eps_4R^2 \bigg]\labell{totR}
\eeqa
 The transformation of this action to the string frame produces only an overall dilaton factor. The couplings of the derivatives of the   dilaton and gravity appears in the Lagrangian \reef{del2}, \reef{2del5}, \reef{ab} and \reef{c} which can   be extended to the   $SL(2,Z)$ invariant forms.  The S-duality invariant form  of the dilaton couplings then includes automatically the appropriate R-R scalar couplings. We have not found the complete B-field couplings and the other R-R couplings.

We have used the generalized curvatures \reef{nonlinear2} to construct the couplings in the even-even and the odd-odd sectors in section 2.  This treats the gravity and the B-field on the same footing. However, since the gravity and the B-field transform totally differently under the S-duality, one expects, in a manifestly S-duality invariant action, the B-field couplings do not appear only in the form of the generalized curvatures. So to include all the B-field couplings in \reef{tot}, one may construct  various couplings in the even-even and the odd-odd sectors for $\prt H$, $H^2$ and $R$ with unknown coefficients.  The S-duality then requires the same couplings for R-R two-form.  
Using the constraints that the action must be invariant under   T-duality,   one can relate them to the couplings of the R-R scalar that we have found in this paper. In this way one may be able to find all the unknown  coefficients. Then using the combination of S- and T-dualities, one may be able to find all other couplings. Similar calculation  has been done in \cite{Garousi:2013lja} for finding various on-shell four-field  couplings. It would be intersting to perform these calculations to find a manifestly T-duality and S-duality invariant action.  After finding such action,  one may use appropriate field redefinitions to rewrite the action in a simpler form, \eg converting some of the couplings involving the Ricci and scalar curvatures to  the couplings involving other massless fields. Of course, the   action then would not be manifestly invariant under the dualities.

We have proposed a prescription for constructing the tensors $t_{2n}$ for any integer $n$. The tensor $t_8$   constructed in \reef{t8} is the same as the symmetric trace prescription given by Tseytlin \cite{Tseytlin:1997csa}, \ie $\Str(t_8F^4)=\tr(t_8F^4)$.  However, our construction for tensor $t_{12}$  is different from the symmetric trace construction of    $F^6$. To clarify this let us construct $t_{12}$ tensor. According to our prescription, we have to first consider the expansion \reef{BI} for  six antisymmetric matrices  at order $F^6$ which is given by the following expression:
\beqa
\frac{1}{8}t_{12}F^6&=&-\frac{15}{8}\Tr(F_1F_2)\Tr(F_3F_4)\Tr(F_5F_6)+10 \Tr(F_1F_2F_3)\Tr(F_4F_5F_6)\nonumber\\
&&+\frac{45}{2} \Tr(F_1F_2)\Tr(F_3F_4F_5F_6)-60 \Tr(F_1F_2F_3F_4F_5F_6)\labell{t12}
\eeqa
where we have used the fact that $\Tr(F_i)=0$. Then we have to replace each term with the average of all independent contractions with identical weight, \ie there are 15 different contractions for the first term, 10 different contractions for the second term, 45 different contractions for the third term and 5!=60 contractions for the last term. This fixes the ordering of the antisymmetric matrices.  For the nonabelian gauge field strength, one has to take the trace over the gauge group generators as well. The symmetric trace prescription, on the other hand, first makes each term symmetric under all permutations of the field strength and then takes the trace over the gauge group generators. 
The $(F^3)^2$ terms  in \reef{t12} are removed by the symmetric trace operator. As a result $\Str(t_{12}F^6)\neq \tr(t_{12}F^6)$. On the other hand,  it is known that the  symmetric trace prescription does not work at the order of six gauge field strengths \cite{Hashimoto:1997gm,Bain:1999hu}. At this order one has to include the appropriate commutators   and the covariant derivatives of the field strengths to have consistency between the effective field theory and string theory results. The $(F^3)^2$ terms produce two commutator terms which are zero for abelian gauge field.% It would be interesting to use the above prescription for the couplings $\tr(t_{12}F^6)$ to calculate the spectrum of open strings at order $\alpha'^6$ for the intersecting D$_2$-branes considered in \cite{Hashimoto:1997gm} and compare them with the sting theory results found in  \cite{Hashimoto:1997gm}. 

We have seen that the S-duality constraint forbids us to have the couplings of three Riemann curvatures in the even-even sector, \ie $a_3=0$ in \reef{consa}, and $m_{20}=0$ in \reef{2del5}. This indicates that the   even-even sector does not produce couplings between  three Riemann curvatures. Moreover, it has been observed in \cite{Stieberger:2009rr} that there is no $R^5$ coupling in the superstring theory either. These couplings, however, may be non-zero in the bosonic string theory. Similar situation appears for the   non-abelian gauge field couplings on D-branes world volume theory, \eg  the coupling of  three gauge field strengths is nonzero  in the bosonic string theory whereas this coupling is zero in the superstring theory.  In that case the symmetric trace prescription for non-abelian Born-Infeld action \cite{Tseytlin:1997csa} removes such odd number of gauge field strengths from the Born-Indeld action. Here also one may speculate that in the superstring theory the couplings in the even-even sector which have odd number of Riemann curvatures are zero.

The S-matrix element of six gravitons  has been studied in \cite{Stieberger:2009rr,Green:2013bza}. The coefficient of the $R^6$ couplings has been found in \cite{Stieberger:2009rr} to be proportional to $\z(5)$. On the other hand, the overall dilaton factor of the sphere-level $R^6$ couplings in the Einstein frame is $e^{-5\phi/2}$. Using the fact that the first term of the weak-expansion of the non-holomorphic Eisenstein series $E_{(5/2)}(\tau,\bar{\tau})$ is $\z(5)e^{-5\phi/2}$, one may extend the sphere-level amplitude to include the one-loop and nonperturbative corrections by extending the dilaton factor to  $E_{(5/2)}(\tau,\bar{\tau})$. One may also use the tensor $t_{12}$ in \reef{t12} to construct the tensorial structure of the effective couplings for six Riemann curvatures in the even-even sector.  There is no coupling in the odd-odd sector in 10 dimensions for obvious reason.   For constant dilaton, then the $SL(2,Z)$ invariant action may   be the following:
\beqa
S&\sim&\int d^{10}x\, E_{(5/2)}(\tau, \bar{\tau})\sqrt{-G} \,t_{12}t_{12}\cR^6
\eeqa
Another possibility for the tensorial structure of the couplings is the symmetric trace prescription for $t_{12}$ which removes the $(F^3)^2$ terms in \reef{t12}.
For non-constant dilatons, one may  add to this action the appropriate couplings involving the Ricci and scalar curvatures by making it to be consistent with the S-duality, as we have done in this paper for the couplings at order $R^4$.  We expect one of the two choices for the tensorial structure of the couplings  to be consistent with the S-duality constraint. The above calculation may  then fix the ambiguity of the $(F^3)^{2}$ terms.

{\bf Acknowledgments}:    This work is supported by Ferdowsi University of Mashhad.


\begin{thebibliography}{99}

%\cite{Kikkawa:1984cp}
\bibitem{Kikkawa:1984cp}
  K.~Kikkawa and M.~Yamasaki,
  %``Casimir Effects in Superstring Theories,''
  Phys.\ Lett.\  B {\bf 149}, 357 (1984).
  %%CITATION = PHLTA,B149,357;%%
\bibitem{TB}  
T. Buscher, Phys. Lett. B  {\bf 194} (1987) 59; B {\bf 201} (1988) 466.

%\cite{Giveon:1994fu}
\bibitem{Giveon:1994fu}
  A.~Giveon, M.~Porrati and E.~Rabinovici,
  %``Target space duality in string theory,''
  Phys.\ Rept.\  {\bf 244}, 77 (1994)
  [arXiv:hep-th/9401139].
  %%CITATION = PRPLC,244,77;%%

%\cite{Alvarez:1994dn}
\bibitem{Alvarez:1994dn}
  E.~Alvarez, L.~Alvarez-Gaume and Y.~Lozano,
  %``An Introduction to T duality in string theory,''
  Nucl.\ Phys.\ Proc.\ Suppl.\  {\bf 41}, 1 (1995)
  [arXiv:hep-th/9410237].
  %%CITATION = NUPHZ,41,1;%%
  
 %\cite{Meessen:1998qm}
\bibitem{Meessen:1998qm}
  P.~Meessen and T.~Ortin,
  %``An Sl(2,Z) multiplet of nine-dimensional type II supergravity theories,''
  Nucl.\ Phys.\  B {\bf 541}, 195 (1999)
  [arXiv:hep-th/9806120].
  %%CITATION = NUPHA,B541,195;%%
  
%\cite{Bergshoeff:1995as}
\bibitem{Bergshoeff:1995as}
  E.~Bergshoeff, C.~M.~Hull and T.~Ortin,
  %``Duality in the type II superstring effective action,''
  Nucl.\ Phys.\  B {\bf 451}, 547 (1995)
  [arXiv:hep-th/9504081].
  %%CITATION = NUPHA,B451,547;%%
%\cite{Bergshoeff:1996ui}
\bibitem{Bergshoeff:1996ui}
  E.~Bergshoeff, M.~de Roo, M.~B.~Green, G.~Papadopoulos and P.~K.~Townsend,
  %``Duality of Type II 7-branes and 8-branes,''
  Nucl.\ Phys.\  B {\bf 470}, 113 (1996)
  [arXiv:hep-th/9601150].
  %%CITATION = NUPHA,B470,113;%%

%\cite{Hassan:1999bv}
\bibitem{Hassan:1999bv}
  S.~F.~Hassan,
  %``T-duality, space-time spinors and R-R fields in curved backgrounds,''   
  Nucl.\ Phys.\  B {\bf 568}, 145 (2000)
  [arXiv:hep-th/9907152].
  %%CITATION = NUPHA,B568,145;%% 	
	 
  
 %\cite{Font:1990gx}
\bibitem{Font:1990gx}
  A.~Font, L.~E.~Ibanez, D.~Lust and F.~Quevedo,
  %``Strong - weak coupling duality and nonperturbative effects in string
  %theory,''
  Phys.\ Lett.\  B {\bf 249}, 35 (1990).
  %%CITATION = PHLTA,B249,35;%%
  %\cite{Rey:1989xj}
\bibitem{Rey:1989xj}
  S.~J.~Rey,
  %``THE CONFINING PHASE OF SUPERSTRINGS AND AXIONIC STRINGS,''
  Phys.\ Rev.\  D {\bf 43}, 526 (1991).
  %%CITATION = PHRVA,D43,526;%%

%\cite{Sen:1994fa}
\bibitem{Sen:1994fa}
  A.~Sen,
  %``Strong - weak coupling duality in four-dimensional string theory,''
  Int.\ J.\ Mod.\ Phys.\  A {\bf 9}, 3707 (1994)
  [arXiv:hep-th/9402002].
  %%CITATION = IMPAE,A9,3707;%%
%\cite{Sen:1994yi}
\bibitem{Sen:1994yi}
  A.~Sen,
  %``Dyon - monopole bound states, selfdual harmonic forms on the multi -
  %monopole moduli space, and SL(2,Z) invariance in string theory,''
  Phys.\ Lett.\  B {\bf 329}, 217 (1994)
  [arXiv:hep-th/9402032].
  %%CITATION = PHLTA,B329,217;%%
	
	%\cite{Schwarz:1993cr}
\bibitem{Schwarz:1993cr}
  J.~H.~Schwarz,
  %``Does string theory have a duality symmetry relating weak and strong
  %coupling?,''
  arXiv:hep-th/9307121.
  %%CITATION = HEP-TH/9307121;%%

%\cite{Hull:1994ys}
\bibitem{Hull:1994ys}
  C.~M.~Hull and P.~K.~Townsend,
  %``Unity of superstring dualities,''
  Nucl.\ Phys.\  B {\bf 438}, 109 (1995)
  [arXiv:hep-th/9410167].
  %%CITATION = NUPHA,B438,109;%% 


  
  %\cite{Garousi:2009dj}
\bibitem{Garousi:2009dj} 
  M.~R.~Garousi,
  %``T-duality of Curvature terms in D-brane actions,''
  JHEP {\bf 1002}, 002 (2010)
  [arXiv:0911.0255 [hep-th]].
  %%CITATION = ARXIV:0911.0255;%%
  %16 citations counted in INSPIRE as of 23 Oct 2013
  
 %\cite{Garousi:2010ki}
\bibitem{Garousi:2010ki} 
  M.~R.~Garousi,
  %``Ramond-Ramond field strength couplings on D-branes,''
  JHEP {\bf 1003}, 126 (2010)
  [arXiv:1002.0903 [hep-th]].
  %%CITATION = ARXIV:1002.0903;%%
  %15 citations counted in INSPIRE as of 23 Oct 2013
  %\cite{Becker:2010ij}
\bibitem{Becker:2010ij} 
  K.~Becker, G.~Guo and D.~Robbins,
  %``Higher Derivative Brane Couplings from T-Duality,''
  JHEP {\bf 1009}, 029 (2010)
  [arXiv:1007.0441 [hep-th]].
  %%CITATION = ARXIV:1007.0441;%%
  %18 citations counted in INSPIRE as of 23 Oct 2013
  %\cite{Garousi:2010rn}
\bibitem{Garousi:2010rn} 
  M.~R.~Garousi,
  %``T-duality of anomalous Chern-Simons couplings,''
  Nucl.\ Phys.\ B {\bf 852}, 320 (2011)
  [arXiv:1007.2118 [hep-th]].
  %%CITATION = ARXIV:1007.2118;%%
  %13 citations counted in INSPIRE as of 23 Oct 2013
	
%\cite{Garousi:2010bm}
\bibitem{Garousi:2010bm} 
  M.~R.~Garousi and M.~Mir,
  %``On RR couplings on D-branes at order $O(\alpha'^2)$,''
  JHEP {\bf 1102}, 008 (2011)
  [arXiv:1012.2747 [hep-th]].
  %%CITATION = ARXIV:1012.2747;%%
  %15 citations counted in INSPIRE as of 23 Oct 2013	
	
 %\cite{Garousi:2011ut}
\bibitem{Garousi:2011ut} 
  M.~R.~Garousi and M.~Mir,
  %``Towards extending the Chern-Simons couplings at order $O(\alpha'^2)$,''
  JHEP {\bf 1105}, 066 (2011)
  [arXiv:1102.5510 [hep-th]].
  %%CITATION = ARXIV:1102.5510;%%
  %10 citations counted in INSPIRE as of 23 Oct 2013 
%\cite{Garousi:2011fc}
\bibitem{Garousi:2011fc} 
  M.~R.~Garousi,
  %``S-duality of D-brane action at order $O(\alpha'^2)$,''
  Phys.\ Lett.\ B {\bf 701}, 465 (2011)
  [arXiv:1103.3121 [hep-th]].
  %%CITATION = ARXIV:1103.3121;%%
  %13 citations counted in INSPIRE as of 23 Oct 2013  
 %\cite{Becker:2011bw}
\bibitem{Becker:2011bw} 
  K.~Becker, G.~-Y.~Guo and D.~Robbins,
  %``Disc amplitudes, picture changing and space-time actions,''
  JHEP {\bf 1201}, 127 (2012)
  [arXiv:1106.3307 [hep-th]].
  %%CITATION = ARXIV:1106.3307;%%
  %7 citations counted in INSPIRE as of 23 Oct 2013 
  %\cite{Becker:2011ar}
\bibitem{Becker:2011ar} 
  K.~Becker, G.~Guo and D.~Robbins,
  %``Four-Derivative Brane Couplings from String Amplitudes,''
  JHEP {\bf 1112}, 050 (2011)
  [arXiv:1110.3831 [hep-th]].
  %%CITATION = ARXIV:1110.3831;%%
  %9 citations counted in INSPIRE as of 23 Oct 2013
  %\cite{Velni:2012sv}
\bibitem{Velni:2012sv} 
  K.~B.~Velni and M.~R.~Garousi,
  %``S-matrix elements from T-duality,''
  Nucl.\ Phys.\ B {\bf 869}, 216 (2013)
  [arXiv:1204.4978 [hep-th]].
  %%CITATION = ARXIV:1204.4978;%%
  %6 citations counted in INSPIRE as of 23 Oct 2013
%\cite{McOrist:2012yc}
\bibitem{McOrist:2012yc} 
  J.~McOrist and S.~Sethi,
  %``M-theory and Type IIA Flux Compactifications,''
  JHEP {\bf 1212}, 122 (2012)
  [arXiv:1208.0261 [hep-th]].
  %%CITATION = ARXIV:1208.0261;%%
  %18 citations counted in INSPIRE as of 26 Oct 2013	
	
%\cite{Kahle:2013kg}
\bibitem{Kahle:2013kg} 
  A.~Kahle and R.~Minasian,
  %``D-brane couplings and Generalised Geometry,''
  arXiv:1301.7238 [hep-th].
  %%CITATION = ARXIV:1301.7238;%%
  %6 citations counted in INSPIRE as of 26 Oct 2013		
	
  
	
 %\cite{Garousi:2013gea}
\bibitem{Garousi:2013gea} 
  M.~R.~Garousi, A.~Ghodsi, T.~Houri and G.~Jafari,
  %``T-duality of D-brane action at order $\alpha'$ in bosonic string theory,''
  JHEP {\bf 1310}, 103 (2013)
  [arXiv:1308.4609 [hep-th]].
  %%CITATION = ARXIV:1308.4609;%% 
	
		
	
	%\cite{Garousi:2012yr}
\bibitem{Garousi:2012yr} 
  M.~R.~Garousi,
  %``T-duality of the Riemann curvature corrections to supergravity,''
  Phys.\ Lett.\ B {\bf 718}, 1481 (2013)
  [arXiv:1208.4459 [hep-th]].
  %%CITATION = ARXIV:1208.4459;%%
  %9 citations counted in INSPIRE as of 23 Oct 2013
	%\cite{Garousi:2012jp}
\bibitem{Garousi:2012jp} 
  M.~R.~Garousi,
  %``Ricci curvature corrections to type II supergravity,''
  Phys.\ Rev.\ D {\bf 87}, 025006 (2013)
  [arXiv:1210.4379 [hep-th]].
  %%CITATION = ARXIV:1210.4379;%%
  %5 citations counted in INSPIRE as of 25 Jun 2013
	
	
	

%\cite{Garousi:2013lja}
\bibitem{Garousi:2013lja} 
  M.~R.~Garousi,
  %``Ramond-Ramond corrections to type II supergravity at order $\alpha'^3$,''
  JHEP {\bf 1306}, 030 (2013)
  [arXiv:1302.7275 [hep-th]].
  %%CITATION = ARXIV:1302.7275;%%
  %1 citations counted in INSPIRE as of 25 Jun 2013
	
	%\cite{Garousi:2013zca}
\bibitem{Garousi:2013zca} 
  M.~R.~Garousi,
  %``Generalized Riemann curvature corrections to type II supergravity,''
  Phys.\ Rev.\ D {\bf 88}, 024033 (2013)
  [arXiv:1303.4034 [hep-th]].
  %%CITATION = ARXIV:1303.4034;%%
  %2 citations counted in INSPIRE as of 23 Oct 2013
	%\cite{Liu:2013dna}
\bibitem{Liu:2013dna} 
  J.~T.~Liu and R.~Minasian,
  %``Higher-derivative couplings in string theory: dualities and the B-field,''
  arXiv:1304.3137 [hep-th].
  %%CITATION = ARXIV:1304.3137;%%
  %2 citations counted in INSPIRE as of 25 Jun 2013
	
	%\cite{Godazgar:2013bja}
\bibitem{Godazgar:2013bja} 
  H.~Godazgar and M.~Godazgar,
  %``Duality completion of higher derivative corrections,''
  JHEP {\bf 1309}, 140 (2013)
  [arXiv:1306.4918 [hep-th]].
  %%CITATION = ARXIV:1306.4918;%%
  %2 citations counted in INSPIRE as of 23 Oct 2013

  
  
	
	%\cite{Garousi:2013tca}
\bibitem{Garousi:2013tca} 
  M.~R.~Garousi,
  %``S-duality invariant dilaton couplings at order $\alpha'^3$,''
 JHEP {\bf 076}, 10 (2013) [arXiv:1306.6851 [hep-th]].
  %%CITATION = ARXIV:1306.6851;%%
%\cite{Maxfield:2013wka}
\bibitem{Maxfield:2013wka} 
  T.~Maxfield, J.~McOrist, D.~Robbins and S.~Sethi,
  %``New Examples of Flux Vacua,''
  arXiv:1309.2577 [hep-th].
  %%CITATION = ARXIV:1309.2577;%%
  %1 citations counted in INSPIRE as of 26 Oct 2013  
	

  


%\cite{Schwarz:1982jn}
\bibitem{Schwarz:1982jn} 
  J.~H.~Schwarz,
  %``Superstring Theory,''
  Phys.\ Rept.\  {\bf 89}, 223 (1982).
  %%CITATION = PRPLC,89,223;%%
  %1081 citations counted in INSPIRE as of 25 May 2013



 %\cite{Gross:1986iv}
\bibitem{Gross:1986iv}
  D.~J.~Gross and E.~Witten,
  %``Superstring Modifications Of Einstein's Equations,''
  Nucl.\ Phys.\  B {\bf 277}, 1 (1986).
  %%CITATION = NUPHA,B277,1;%%
  
 
%\cite{Grisaru:1986vi}
\bibitem{Grisaru:1986vi} 
  M.~T.~Grisaru and D.~Zanon,
  %``Sigma Model Superstring Corrections To The Einstein-hilbert Action,''  
  Phys.\ Lett.\ B {\bf 177}, 347 (1986).  
  %%CITATION = PHLTA,B177,347;%%
 %\cite{Freeman:1986zh}
\bibitem{Freeman:1986zh} 
  M.~D.~Freeman, C.~N.~Pope, M.~F.~Sohnius and K.~S.~Stelle,
  %``Higher Order Sigma Model Counterterms And The Effective Action For Superstrings,'' 
   Phys.\ Lett.\ B {\bf 178}, 199 (1986).  
   %%CITATION = PHLTA,B178,199;%%

  
  %\cite{Zumino:1985dp}
\bibitem{Zumino:1985dp} 
  B.~Zumino,
  %``Gravity Theories in More Than Four-Dimensions,''  
  Phys.\ Rept.\  {\bf 137}, 109 (1986).  
  %%CITATION = PRPLC,137,109;%%  
	

 
%\cite{Green:1997tv}
\bibitem{Green:1997tv}
  M.~B.~Green and M.~Gutperle,
  %``Effects of D instantons,''
  Nucl.\ Phys.\  B {\bf 498}, 195 (1997)
  [arXiv:hep-th/9701093].
  %%CITATION = NUPHA,B498,195;%%
%\cite{Green:1997di}
\bibitem{Green:1997di}
  M.~B.~Green and P.~Vanhove,
  %``D instantons, strings and M theory,''
  Phys.\ Lett.\  B {\bf 408}, 122 (1997)
  [arXiv:hep-th/9704145].
  %%CITATION = PHLTA,B408,122;%%
%\cite{Green:1997as}
\bibitem{Green:1997as}
  M.~B.~Green, M.~Gutperle and P.~Vanhove,
  %``One loop in eleven-dimensions,''
  Phys.\ Lett.\  B {\bf 409}, 177 (1997)
  [arXiv:hep-th/9706175].
  %%CITATION = PHLTA,B409,177;%%
%\cite{Kiritsis:1997em}
\bibitem{Kiritsis:1997em}
  E.~Kiritsis and B.~Pioline,
  %``On R**4 threshold corrections in IIb string theory and (p, q) string
  %instantons,''
  Nucl.\ Phys.\  B {\bf 508}, 509 (1997)
  [arXiv:hep-th/9707018].
  %%CITATION = NUPHA,B508,509;%%
%\cite{Green:1997me}
\bibitem{Green:1997me}
  M.~B.~Green, M.~Gutperle and H.~h.~Kwon,
  %``Sixteen fermion and related terms in M theory on T**2,''
  Phys.\ Lett.\  B {\bf 421}, 149 (1998)
  [arXiv:hep-th/9710151].
  %%CITATION = PHLTA,B421,149;%%
%\cite{Pioline:1998mn}
\bibitem{Pioline:1998mn}
  B.~Pioline,
  %``A Note on nonperturbative R**4 couplings,''
  Phys.\ Lett.\  B {\bf 431}, 73 (1998)
  [arXiv:hep-th/9804023].
  %%CITATION = PHLTA,B431,73;%%
  %\cite{Green:1998by}
\bibitem{Green:1998by} 
  M.~B.~Green and S.~Sethi,
  %``Supersymmetry constraints on type IIB supergravity,''
  Phys.\ Rev.\ D {\bf 59}, 046006 (1999)
  [hep-th/9808061].
  %%CITATION = HEP-TH/9808061;%%
  %148 citations counted in INSPIRE as of 03 Jul 2013
  
  
%\cite{Green:1999pu}
\bibitem{Green:1999pu}
  M.~B.~Green, H.~h.~Kwon and P.~Vanhove,
  %``Two loops in eleven-dimensions,''
  Phys.\ Rev.\  D {\bf 61}, 104010 (2000)
  [arXiv:hep-th/9910055].
  %%CITATION = PHRVA,D61,104010;%%
%\cite{Obers:1999es}
\bibitem{Obers:1999es}
  N.~A.~Obers and B.~Pioline,
  %``Eisenstein series in string theory,''
  Class.\ Quant.\ Grav.\  {\bf 17}, 1215 (2000)
  [arXiv:hep-th/9910115].
  %%CITATION = CQGRD,17,1215;%%

%\cite{Sinha:2002zr}
\bibitem{Sinha:2002zr}
  A.~Sinha,
  %``The G(hat)**4 lambda**16 term in IIB supergravity,''
  JHEP {\bf 0208}, 017 (2002)
  [arXiv:hep-th/0207070].
  %%CITATION = JHEPA,0208,017;%%
%\cite{Berkovits:2004px}
\bibitem{Berkovits:2004px}
  N.~Berkovits,
  %``Multiloop amplitudes and vanishing theorems using the pure spinor formalism
  %for the superstring,''
  JHEP {\bf 0409}, 047 (2004)
  [arXiv:hep-th/0406055].
  %%CITATION = JHEPA,0409,047;%%
%\cite{D'Hoker:2005jc}
\bibitem{D'Hoker:2005jc}
  E.~D'Hoker and D.~H.~Phong,
  %``Two-loop superstrings VI: Non-renormalization theorems and the 4-point
  %function,''
  Nucl.\ Phys.\  B {\bf 715}, 3 (2005)
  [arXiv:hep-th/0501197].
  %%CITATION = NUPHA,B715,3;%%
%\cite{D'Hoker:2005ht}
\bibitem{D'Hoker:2005ht}
  E.~D'Hoker, M.~Gutperle and D.~H.~Phong,
  %``Two-loop superstrings and S-duality,''
  Nucl.\ Phys.\  B {\bf 722}, 81 (2005)
  [arXiv:hep-th/0503180].
  %%CITATION = NUPHA,B722,81;%%
  %\cite{Matone:2005vm}
\bibitem{Matone:2005vm} 
  M.~Matone and R.~Volpato,
  %``Higher genus superstring amplitudes from the geometry of moduli space,''
  Nucl.\ Phys.\ B {\bf 732}, 321 (2006)
  [hep-th/0506231].
  %%CITATION = HEP-TH/0506231;%%
  %28 citations counted in INSPIRE as of 06 Mar 2013
  
%\cite{Green:2005ba}
\bibitem{Green:2005ba}
  M.~B.~Green and P.~Vanhove,
  %``Duality and higher derivative terms in M theory,''
  JHEP {\bf 0601}, 093 (2006)
  [arXiv:hep-th/0510027].
  %%CITATION = JHEPA,0601,093;%%
%\cite{Green:2006gt}
\bibitem{Green:2006gt}
  M.~B.~Green, J.~G.~Russo and P.~Vanhove,
  %``Non-renormalisation conditions in type II string theory and maximal
  %supergravity,''
  JHEP {\bf 0702}, 099 (2007)
  [arXiv:hep-th/0610299].
  %%CITATION = JHEPA,0702,099;%%
%\cite{Basu:2007ru}
\bibitem{Basu:2007ru}
  A.~Basu,
  %``The D**4 R**4 term in type IIB string theory on T**2 and U-duality,''
  Phys.\ Rev.\  D {\bf 77}, 106003 (2008)
  [arXiv:0708.2950 [hep-th]].
  %%CITATION = PHRVA,D77,106003;%%
%\cite{Basu:2007ck}
\bibitem{Basu:2007ck}
  A.~Basu,
  %``The D**6 R**4 term in type IIB string theory on T**2 and U-duality,''
  Phys.\ Rev.\  D {\bf 77}, 106004 (2008)
  [arXiv:0712.1252 [hep-th]].
  %%CITATION = PHRVA,D77,106004;%%


%\cite{Richards:2008jg}
\bibitem{Richards:2008jg} 
  D.~M.~Richards,
  %``The One-Loop Five-Graviton Amplitude and the Effective Action,''
  JHEP {\bf 0810}, 042 (2008)
  [arXiv:0807.2421 [hep-th]].
  %%CITATION = ARXIV:0807.2421;%%
  %10 citations counted in INSPIRE as of 25 Jun 2013


	
	
%\cite{Vafa:1995fj}
\bibitem{Vafa:1995fj} 
  C.~Vafa and E.~Witten,
  %``A One loop test of string duality,''
  Nucl.\ Phys.\ B {\bf 447}, 261 (1995)
  [hep-th/9505053].
  %%CITATION = HEP-TH/9505053;%%
  %200 citations counted in INSPIRE as of 23 Oct 2013	

%\cite{Duff:1995wd}
\bibitem{Duff:1995wd} 
  M.~J.~Duff, J.~T.~Liu and R.~Minasian,
  %``Eleven-dimensional origin of string-string duality: A One loop test,''
  Nucl.\ Phys.\ B {\bf 452}, 261 (1995)
  [hep-th/9506126].
  %%CITATION = HEP-TH/9506126;%%
  %282 citations counted in INSPIRE as of 23 Oct 2013
	
%\cite{Gross:1986mw}
\bibitem{Gross:1986mw} 
  D.~J.~Gross and J.~H.~Sloan,
  %``The Quartic Effective Action for the Heterotic String,''  
  Nucl.\ Phys.\ B {\bf 291}, 41 (1987).  
  %%CITATION = NUPHA,B291,41;%%

	
%\cite{Richards:2008sa}
\bibitem{Richards:2008sa} 
  D.~M.~Richards,
  %``The One-Loop H**2 R**3 and H**2(Delta H)R-2 Terms in the Effective Action,''
  JHEP {\bf 0810}, 043 (2008)
  [arXiv:0807.3453 [hep-th]].
  %%CITATION = ARXIV:0807.3453;%%
  %8 citations counted in INSPIRE as of 23 Oct 2013	

%\cite{Garousi:2011we}
\bibitem{Garousi:2011we} 
  M.~R.~Garousi,
  %``S-duality of S-matrix,''
  JHEP {\bf 1111}, 016 (2011)
  [arXiv:1106.1714 [hep-th]].
  %%CITATION = ARXIV:1106.1714;%%
  %10 citations counted in INSPIRE as of 26 Oct 2013
%\cite{Garousi:2011vs}
\bibitem{Garousi:2011vs} 
  M.~R.~Garousi,
  %``On S-duality of D$_3$-brane S-matrix,''
  Phys.\ Rev.\ D {\bf 84}, 126019 (2011)
  [arXiv:1108.4782 [hep-th]].
  %%CITATION = ARXIV:1108.4782;%%
  %7 citations counted in INSPIRE as of 26 Oct 2013	
	%\cite{Garousi:2011jh}
\bibitem{Garousi:2011jh} 
  M.~R.~Garousi,
  %``S-duality of color-ordered amplitudes,''
  Nucl.\ Phys.\ B {\bf 862}, 107 (2012)
  [arXiv:1109.5555 [hep-th]].
  %%CITATION = ARXIV:1109.5555;%%
  %3 citations counted in INSPIRE as of 26 Oct 2013
%\cite{Garousi:2012gh}
\bibitem{Garousi:2012gh} 
  M.~R.~Garousi,
  %``Tree-level S-matrix elements from S-duality,''
  JHEP {\bf 1204}, 140 (2012)
  [arXiv:1201.2556 [hep-th]].
  %%CITATION = ARXIV:1201.2556;%%
  %5 citations counted in INSPIRE as of 26 Oct 2013	
	
	

%\cite{Peeters:2001ub}
\bibitem{Peeters:2001ub} 
  K.~Peeters, P.~Vanhove and A.~Westerberg,
  %``Chiral splitting and world sheet gravitinos in higher derivative string amplitudes,''
  Class.\ Quant.\ Grav.\  {\bf 19}, 2699 (2002)
  [hep-th/0112157].
  %%CITATION = HEP-TH/0112157;%%
  %34 citations counted in INSPIRE as of 23 Oct 2013

%\cite{Tseytlin:1996it}
\bibitem{Tseytlin:1996it}
  A.~A.~Tseytlin,
  %``Selfduality of Born-Infeld action and Dirichlet three-brane of type IIB
  %superstring theory,''
  Nucl.\ Phys.\  B {\bf 469}, 51 (1996)
  [arXiv:hep-th/9602064].
  %%CITATION = NUPHA,B469,51;%%
%\cite{Green:1996qg}
\bibitem{Green:1996qg}
  M.~B.~Green and M.~Gutperle,
  %``Comments on three-branes,''                     
  Phys.\ Lett.\  B {\bf 377}, 28 (1996)
  [arXiv:hep-th/9602077].
  %%CITATION = PHLTA,B377,28;%% 
 %\cite{Gibbons:1995ap}
\bibitem{Gibbons:1995ap}
  G.~W.~Gibbons and D.~A.~Rasheed,
  %``Sl(2,R) invariance of nonlinear electrodynamics coupled to an axion and a
  %dilaton,''
  Phys.\ Lett.\  B {\bf 365}, 46 (1996)
  [arXiv:hep-th/9509141].
  %%CITATION = PHLTA,B365,46;%%   
%\cite{Kawai:1985xq}
\bibitem{Kawai:1985xq} 
  H.~Kawai, D.~C.~Lewellen and S.~H.~H.~Tye,
  %``A Relation Between Tree Amplitudes of Closed and Open Strings,''
  Nucl.\ Phys.\ B {\bf 269}, 1 (1986).
  %%CITATION = NUPHA,B269,1;%%
  %358 citations counted in INSPIRE as of 23 Oct 2013

%\cite{Tseytlin:1997csa}
\bibitem{Tseytlin:1997csa} 
  A.~A.~Tseytlin,
  %``On nonAbelian generalization of Born-Infeld action in string theory,''
  Nucl.\ Phys.\ B {\bf 501}, 41 (1997)
  [hep-th/9701125].
  %%CITATION = HEP-TH/9701125;%%
  %450 citations counted in INSPIRE as of 23 Oct 2013

%\cite{Hashimoto:1997gm}
\bibitem{Hashimoto:1997gm} 
  A.~Hashimoto and W.~Taylor,
  %``Fluctuation spectra of tilted and intersecting D-branes from the Born-Infeld action,''
  Nucl.\ Phys.\ B {\bf 503}, 193 (1997)
  [hep-th/9703217].
  %%CITATION = HEP-TH/9703217;%%
  %177 citations counted in INSPIRE as of 23 Oct 2013
	%\cite{Bain:1999hu}
\bibitem{Bain:1999hu} 
  P.~Bain,
  %``On the nonAbelian Born-Infeld action,''
  hep-th/9909154.
  %%CITATION = HEP-TH/9909154;%%
  %39 citations counted in INSPIRE as of 23 Oct 2013
	%\cite{Stieberger:2009rr}
\bibitem{Stieberger:2009rr} 
  S.~Stieberger,
  %``Constraints on Tree-Level Higher Order Gravitational Couplings in Superstring Theory,''
  Phys.\ Rev.\ Lett.\  {\bf 106}, 111601 (2011)
  [arXiv:0910.0180 [hep-th]].
  %%CITATION = ARXIV:0910.0180;%%
  %19 citations counted in INSPIRE as of 23 Oct 2013
	
	%\cite{Green:2013bza}
\bibitem{Green:2013bza} 
  M.~B.~Green, C.~R.~Mafra and O.~Schlotterer,
  %``Multiparticle one-loop amplitudes and S-duality in closed superstring theory,''
  arXiv:1307.3534 [hep-th].
  %%CITATION = ARXIV:1307.3534;%%
  %3 citations counted in INSPIRE as of 23 Oct 2013
	




 
  
 
\end{thebibliography}
\end{document}